\documentclass[fleqn,usenatbib]{mnras}

\usepackage{newtxtext,newtxmath}

\usepackage[T1]{fontenc}
\usepackage{ae,aecompl}
\usepackage [latin1]{inputenc}

\usepackage{graphicx}   

\usepackage{color}
\usepackage{multirow}
\usepackage{bm}

\newcommand\be{\begin{equation}}
\newcommand\en{\end{equation}}
\newcommand\msun{M_{\odot}}

\newcommand\mdot{\dot{M}}

\newcommand{\lgg}{\log(g)}
\newcommand{\lga}{\log({\rm Age})}

\newcommand{\g}{{\it Gaia }}

\newcommand{\gethree}{{\it Gaia} EDR3 }

\newcommand\gj{G_{j}^{k}}
\newcommand\sgj{\sigma_{j}^{k}}
\newcommand\pgj{\widehat{G}_{j}^{k}}

\newcommand\mj{M_{j}^{k}}
\newcommand\avj{A_{V_j}}

\newcommand\mi{\mathcal{M}_{i}^{k}}

\newcommand\pav{p_{i}(A_{V})}
\newcommand\av{A_{V}}

\newcommand{\mlc}{\multicolumn}

\usepackage[dvipsnames]{xcolor}

\defcitealias{alzate21}{AL21}
\defcitealias{mck19}{MK19}
\defcitealias{kounkel18}{K18}

\newcommand{\kss}{\citetalias{kounkel18}}

\title[Star formation in Orion A] {Constraints on star formation in Orion A from Gaia}
  \author[Alzate et al.]
      {Jairo A. Alzate$^{1}$\thanks{E-mail: jalzate@cefca.es}, Gustavo Bruzual$^{2}$, Marina Kounkel$^{3}$, Gladis Magris$^{4}$, Lee Hartmann$^{5}$,
      Nuria Calvet$^{5}$,
      \newauthor
      and Lyra Cao$^{6}$
\\
$^{1}$Instituto de Astrof\'isica \'Optica y Electr\'onica, INAOE, Puebla, Puebla, C.P. 72840, M\'exico \\
$^{2}$Instituto de Radioastronom\'ia y Astrof\'isica, UNAM, Campus Morelia, Michoac\'an, C.P. 58089, M\'exico \\
$^{3}$Dept. of Physics and Astronomy, Vanderbilt University, Nashville, TN 37235, USA\\
$^{4}$Centro de Investigaciones de Astronom\'ia, M\'erida, Venezuela\\
$^{5}$Department of Astronomy, University of Michigan,  500
           Church Street, Ann Arbor, MI 48105, USA \\
$^{6}$Department of Astronomy, The Ohio State University, Columbus, OH 43210, USA\\
}
\date{Accepted XXX. Received YYY; in original form ZZZ}

\pubyear{2022}

\begin{document}
\label{firstpage}
\pagerange{\pageref{firstpage}--\pageref{lastpage}} \maketitle

\begin{abstract}
We develop statistical methods within a Bayesian framework to infer the star formation history from photometric surveys of pre-main sequence populations. Our procedures include correcting for
biases due to extinction in magnitude-limited surveys, and using distributions from subsets of stars with individual extinction measurements.
We also make modest corrections for unresolved
binaries. We apply our methods to samples of populations with Gaia photometry in the Orion A molecular cloud. Using two well-established
sets of evolutionary tracks, we find that, although our sample is incomplete at youngest ages due to extinction, star formation has proceeded in Orion A at a relatively constant
rate between ages of about 0.3 and 5 Myr, in contrast to other studies suggesting multiple epochs of star formation. Similar results are obtained for a set of tracks that attempt to take the effects of strong magnetic fields into account. 
We also find no evidence for a well-constrained ``birthline'' that would result from low-mass stars appearing first along the deuterium-burning main sequence, especially using the magnetic evolutionary tracks. While our methods have been developed to deal with Gaia data, they may be useful for analyzing other photometric
surveys of star-forming regions.
\end{abstract}


\section{Introduction}

The wealth of data from the Gaia mission
provides a new means to study the star formation history (SFH) of nearby molecular clouds and associations. It also presents an opportunity to
systematically assess the initial properties of young stars after the end of the
protostellar phase, potentially yielding insights into the star formation process.
Gaia parallaxes and proper motions are crucial in providing cleaner membership
samples in stellar associations. The systematic optical photometry from
Gaia is also useful because colour-magnitude diagrams (CMDs) may have advantages over HR diagrams, because standard reddening vectors tend to shift late-G and KM stars roughly along isochrones (while extinction corrections in the HR diagram change only luminosities, given effective temperatures or spectral types).
In addition, it is also more straightforward to make statistical
corrections for unresolved binary stars (UBS) in CMDs than in HR diagrams.

The Orion region is an obvious choice for studies of pre-main sequence
as it comprises the nearest set of well-populated associations.
Several studies of the SFH in the
Orion region have been made with extensive use of Gaia data.
For example, 
\citet[][K18 hereafter]{kounkel18} and \citet{zari19}
studied the overall structure and distribution of
several regions of Orion in addition to the Orion A cloud, and characterized their average ages, while 
\citet{grosschedl2018} focused on the details of
the structure of the Orion A cloud.
Distinctively, \citet[][B17 hereafter]{beccari17} and \citet[][J19 hereafter]{jerabkova19}
derive a detailed SFH of a region around the Orion Nebula Cluster (ONC),
which appears to span $\pm 2^{\circ}$ north-south and nearly the same distance east-west (see figure 3 in B17)
and it is spatially consistent with the Northern part of the Orion A cloud.\footnote{While B17 and J19 describe their
study as that of the ONC, the region studied
is much larger than that typically ascribed to the
ONC, and includes the L1641 and OMC2/3 regions \citep[see, e.g. figures 1 and 14 in][]{megeath12}.}.
They argued that star formation in the area has been comprised of three very distinct star-forming episodes, based on CMDs from OmegaCam and Gaia photometry. One surprising aspect of these studies is that the multiple sequences in the CMDs correspond to the positions that would result from unresolved binary and triple systems; but B17 and J19 conclude that explaining the data this way would require
implausible distributions of mass ratios for the multiple systems.

Here we focus on the Orion A star-forming cloud, dominated by the ONC, as the site of the major population of the youngest stars, with substantial ongoing protostar formation \citep[e.g.,][]{megeath12}. 
Nevertheless, the sample used in this work shows a need for stars in the ONC center, where \g's completeness drop rapidly. In such case, our results may be more informative for stars out of the innermost nebula region (Sec.\,\ref{sec:sample}).
We build on the methods developed by \citet[][AL21 hereafter]{alzate21} to analyze the SFH and the possible limits on birthlines in Orion A, making statistical corrections for extinction and UBS to more accurately understand the biases they imposed on the understanding of the SFH in previous studies. Our analysis suggests a roughly constant rate of star formation between about 0.3 and 5 Myr rather than well-defined episodes of star formation. We also show that there is little evidence for a well-defined `birthline'', a locus in the CMD (or HR diagram) where stars are thought to appear at the end of the protostellar infall phase; this suggests considerable variability in the initial conditions of forming stars. Finally, the methodology developed in this work may be useful for systematic studies of other star-forming regions using Gaia data.

\section{Member samples} \label{sec:sample}

\begin{figure*}
\begin{center}
    \includegraphics[width=\textwidth]{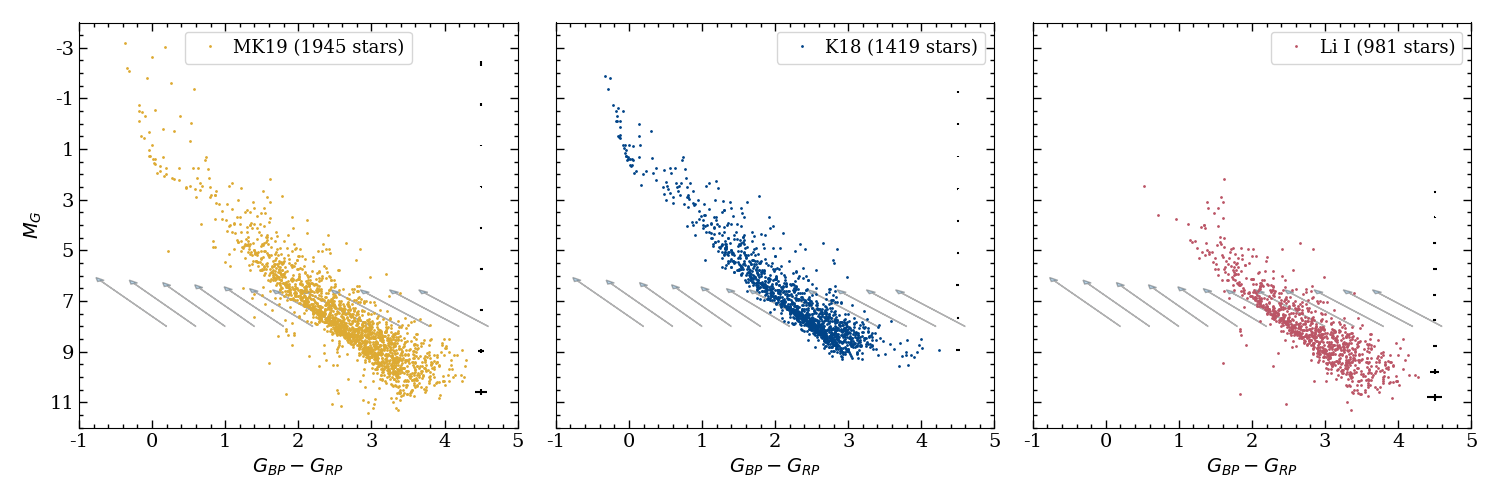}
    \caption{CMD of the cleaned MK19, K18 and Li I samples. The line of arrows show the dereddening vector for the broad band \gethree extinction. The length and the direction of the vectors change as function of the color according to Eq.\,(\ref{eq:hk}) for $\av=2$ mag. The photometric errors are shown in each panel.
    \label{fig:CMD_samples_2}}
\end{center}
\end{figure*}

\begin{figure}
\begin{center}
    \includegraphics[width=\columnwidth]{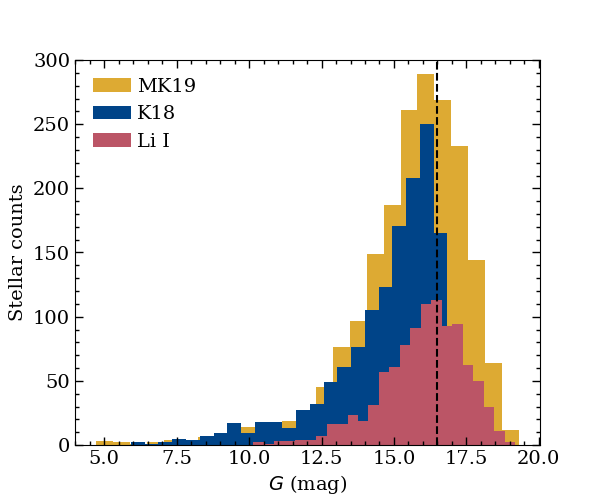}
    \caption{ Stellar counts of the MK19, K18 and Li I samples. The number of stars per bin does grow similarly up to $G_{\rm lim}=16.5$ mag, near to the mode of the
    histograms. The incompleteness in the counts for fainter stars are different for the three samples.
    \label{fig:hist_g}}
\end{center}
\end{figure}

\begin{figure}
\begin{center}
    \includegraphics[width=0.49\textwidth]{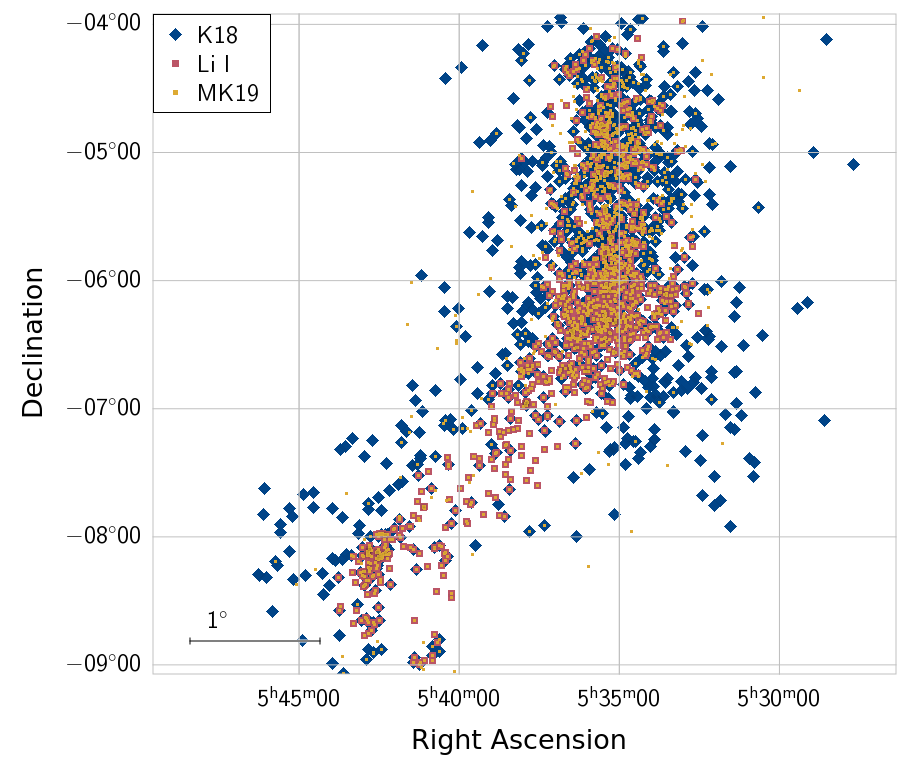}
    \caption{\textbf{Distribution in the sky of the three samples in Fig.\,\ref{fig:CMD_samples_2}.
    \label{fig:map}}
    }
\end{center}
\end{figure}

We assembled multiple catalogs of members of Orion A. One catalog that we use is from \citetalias{kounkel18}, consisting of 1970 stars, is derived through clustering of Gaia DR2 \citep{gaia-collaboration2018} astrometry and APOGEE \citep{ahumada2020} radial velocities. This sample extends down only to $H$\,<\,13 mag ($G$\,$\sim$\,17 mag) due to the bright limit of APOGEE. Individual $A_V$ measurements derived through spectral energy distribution fitting, was also made available as a part of this catalog. 

We also assembled a catalog of sources for which optical spectra have been obtained to observe Li I 6707.7 \AA\ absorption. As Li I depletes rapidly in young convective stars, the presence of this spectral feature is a robust test of youth, allowing to generate a very clean catalog of members \citep[e.g.,][]{briceno1997}. This census consists of 1465 stars with equivalent width EqW$_{\mathrm{Li\ I}}>0.1$\AA, assembled from \citet{sicilia-aguilar2005}, \citet{hsu2012}, \citet{fang2013}, \citet{kounkel2016}, \citet{fang2017}, and \citet{kounkel2017a}. This census is somewhat inhomogeneous, due to targeting strategies utilized by all these surveys, but it allows to verify the underlying age distribution is not affected by contaminants. This catalog extends to lower mass stars than the census from \citetalias{kounkel18}, extending to fainter magnitudes by more than 1 mag as the limit. However, this catalog lacks stars earlier than G-type, as Li I cannot be detected in them.

The main catalog that we used to infer the
Orion A SFH was that assembled by \citet[][MK19 hereafter]{mck19} using a variety of tracers. It consists of 5988 stars in total, with estimated contamination rate of $\sim$1\%. 
Many of these stars are not detectable in the Gaia database. Typically these stars are heavily extincted due to their extreme youth; only 4758 sources can be cross-matched to Gaia EDR3 \citep{gaia-collaboration2021}. Furthermore, Gaia photometry near the Trapezium is strongly affected by nebulosity, making it difficult to compare them to the isochrones. We use the criteria described in the equation C.2 in \citet{lindegren2018} to select the sources with clean photometry, ending with a sample consisting of 3012 stars.

From an examination of the $(G_{BP}-G)$\,vs.\,$G-G_{RP}$ diagram we conclude that it is necessary to restrict the corrected \texttt{phot$\_$bp$\_$rp$\_$excess$\_$factor} \citep{riello2021} to values less than 0.1, in order to build samples with negligible contaminating light from neighboring sources. Additionally, stars with high probability to have spurious parallaxes were rejected using the astrometric fidelity parameter proposed by \citet{rybizki2022}, thus we selected stars with fidelity values larger than 0.5. Finally, we considered that stars with distance outside the interval $[300,500]$ pc are probably foreground/background contaminants. This fairly large range is adopted to accommodate members with large parallax errors due to nebulosity and crowding. 

Fig.\,\ref{fig:CMD_samples_2} shows the CMD of the \citetalias{mck19}, K18 and Li I samples with their final number of stars after filtering for flux excess, fidelity of the parallax, and distance.
Fig.\,\ref{fig:hist_g} displays the corresponding stellar counts, which are essential to roughly estimate the completeness limit of the samples and to avoid bias in the inferred star formation rate (Sec.\,\ref{sec:marg_post} and \ref{sec:corr_ext}).
We use the K18 sample, which has the extinction measurements, to build a statistical method for correcting the \citetalias{mck19} sample for
extinction, as described later. The Li I catalog is used supplementary. Since stars listed in this catalog are members with high confidence, we use it to perform internal checks that confirm that our results are robust against sample definition.

These catalogs cover the entire Orion A molecular cloud. Fig.\,\ref{fig:map} shows their position in the sky in equatorial coordinates. The three samples show significant overlap and offer extensive coverage of the region, except in the zone around $(\alpha,\delta)=(05^{\rm h}\ 35^{\rm m}, -5^{\circ} 23')$, strongly affected by the nebula. Since we know the parallax of each source, the distance distribution of Orion A is used in our model, including those stars in the southern region.

\section{Inferring the star formation history}\label{sec:inference}

\begin{table}
\begin{center}
 \caption{\label{tab:model_param}Variables entering the hierarchical model.$^a$
 }
\begin{tabular}{llcl}
 \hline
 Hyper-parameter                   &             $\bm{a}$              &        &    Stellar fraction vector           \\
 \hline
 \multirow{4}{*}{Parameters ($\bm{\beta}$)}       &              $r_{j}$              &   pc   &    Heliocentric distance             \\
                                   &            $M_{j}^{k}$            &   mag  &    Absolute magnitude                \\
                                   &            $\varpi_{{\rm true},j}$ &   mas  &    True parallax                     \\
                                   &            $G_{{\rm true},j}^{k}$ &   mag  &    True apparent magnitude           \\
 \hline
 \multirow{4}{*}{Data ($\bm{d}$)}             &            $\varpi_{j}$           &   mas  &    Observed parallax                 \\
                                   &           $e_{\varpi,j}$          &   mas  &    Parallax error                    \\
                                   &            $G_{j}^{k}$            &   mag  &    Observed apparent magnitude       \\
                                   &            $e_{j}^{k}$            &   mag  &    Apparent magnitude error          \\
 \hline
 \multirow{4}{*}{Fixed distributions} &             $p_{i}(A_{V})$            &   mas  &    Visual extinction PDF                 \\
                                   &        $\mathcal{M}_{i}^{k}$      &   mag  &    Isochrone absolute magnitude      \\
                                   &          $\sigma_{i}^{k}$         &   mag  &    Isochrone broadening       \\
                                   &              $\phi(m)$            &        &    Initial mass function             \\
 \hline
 \multicolumn{4}{l}{$^a$Subscripts $i$~=~$1,...,N_{\rm I}$, $j$~=~$1,...,N_{D}$ indicate the $i^{th}$ isochrone and}\\
 \multicolumn{4}{l}{the $j^{th}$ star, respectively, where $N_{\rm I}$ is the number of selected isochrones}\\
 \multicolumn{4}{l}{and $N_{D}$ the number of observed stars. Superscript $k$~=~$1,2,3$ refers to the}\\
 \multicolumn{4}{l}{broad (330-1050) nm $G$, blue (330-680) nm $G_{BP}$ and red (630-1050) nm}\\
 \multicolumn{4}{l}{$G_{RP}$ {\it Gaia} photometric bands, respectively.}\\
\end{tabular}
\end{center}
\end{table}

The CMDs of resolved stellar populations contain information about when the stars were born and by how much they have evolved, i.e., about the SFH of the stellar system \citep{dolphin2002,verg2002,aparicio2009,daltio2021}. Combining stellar evolution models and stellar spectral energy distributions we can compute the position at age $t$ of a star of mass $m$ and metallicity $Z$ in any CMD. Pre-main sequence stars in regions such as Orion A generally cannot be characterized by a single isochrone, as their intrinsic age spread of a few Myr is apparent in the CMD, in contrast to their more evolved counterparts. To overcome this limitation, in this paper we follow a statistical approach based on the assumption that the CMD of a stellar population results from a combination of multiple differently populated isochrones \citep{small2013}. The SFH then follows from the statistical weights assigned to the different isochrones. These weights can be inferred using, for example, either maximum likelihood methods \citep{hernandez2008,small2013} or Bayesian techniques (\citetalias{alzate21}).  In this section we extend the \citetalias{alzate21} hierarchical statistical model to study young stellar populations, highly affected by dust, and to allow for the presence of UBS.

Following \citetalias{alzate21}, we define a vector $\bm{a}=(a_{i=1,2,..,N_{I}})$, where $a_{i}$ is the fractional number of stars assigned to the $i$-th isochrone, such that $a_{i}\geq 0$ and $\sum a_{i} = 1$. The posterior probability distribution function (pdf) of $\bm{a}$ and $\bm{\beta}$, given data $\bm{d}$, is stated by the Bayes theorem
\begin{equation}\label{eq:bayes}
    p(\bm{a}, \bm{\beta} \vert {\bm d}) = p(\bm{a}) \prod_{j=1}^{N_{D}} \ \frac{S(d_{j})\mathcal{L}(d_{j} \vert \beta_{j}) \ p(\beta_{j}\vert \bm{a})}{\ell(\bm{a},S)}.
\end{equation}
Here $p$ refers to the pdfs involved in the theorem, $\mathcal{L}$ is the likelihood function, and $N_{D}$ is the number of observed stars. The notation $p(x\vert y)$ is read as the conditional pdf of $x$ given $y$. Random variables, data and priors in Eq.~(\ref{eq:bayes}) are listed in Table\,\ref{tab:model_param}. In our case, the data $\bm{d}$ are \g EDR3 photometric magnitudes and parallaxes, and the $\bm{\beta}$ parameters correspond to individual properties of the stars, the magnitude $\mj$ and the extinction $\avj$, predicted by the stellar evolution model and the extinction distribution, respectively. The indices $i,j,k$ and the numbers $N_I$ and $N_D$ are defined in the note to Table~\ref{tab:model_param}. The empirical distribution of extinction by dust in the $V$ band for the Orion A stars, $\pav$, the initial mass function (IMF), $\phi(m)$, and the absolute magnitude $\mi(m)$ of a star of mass $m$ for a set of isochrones are fixed distributions, selected according to our previous knowledge of the target population. The function $S(d_{j})$ describes the completeness of the sample.

\subsection{Likelihood and prior pdf's}

$\mathcal{L}(d_{j} \vert \beta_{j})$, the likelihood distribution function, assumes that the observed apparent magnitudes, $G,G_{\rm BP},G_{\rm RP}=G^{1},G^{2},G^{3}$, are random measurements of the predicted magnitudes, $\widehat{G},\widehat{G}_{\rm BP},\widehat{G}_{\rm RP}$, with standard deviation $\sigma_{G}, \sigma_{G_{\rm BP}}, \sigma_{G_{\rm RP}}$, respectively. The \g photometric measurements are well approximated by a normal distribution
\begin{equation}\label{eq:normal}
    \mathcal{L}(\gj \vert \mj, \avj) = \frac{1}{\sqrt{2\pi \sgj}}\exp\left[ -\frac{\left(\gj-\pgj\left(\mj,\avj\right) \right)^2}{2{\sgj}^2} \right],
\end{equation}
where 
\begin{equation}\label{eq:pgj}
    \pgj = \mj + f(\varpi_j) + h^{k}\avj
\end{equation}
and
\begin{equation}\label{eq:hk}
    h^{k}=A^{k}/\av
\end{equation}
is the ratio between the $k^{th}$ band and the visual band extinction. 
Due to the broad width of the \g photometric bands, $h^{k}$ changes depending on the effective temperature ($T_{\rm eff}$) and reddening of the observed star \citep{jordi2010}. In consequence, the ratio $h^{k}$~=~$A^{k}/\av$ depends on the intrinsic colour of the star, $(G_{BP}-G_{RP})_{0}$, and its visual extinction $\av$ (Fig.\,\ref{fig:hk}). For this reason, we add the extinction $A^{k}$ to the isochrone absolute magnitude in Eq.~(\ref{eq:pgj}), instead of correcting the data. Given that 99\% of the stars in the three samples defined in Sec.\,\ref{sec:sample} have relative parallax error below 10\%, our statistical model is simplified by approximating the distance likelihood function as a Dirac's $\delta$ function and the distance modulus as
\begin{equation}\label{eq:dm}
    f(\varpi)=-5\log(\varpi/{\rm arcsec})-5,
\end{equation}
with no need to specify a prior pdf for the distance.
We assume that the individual parameters are statistically independent, and that their prior probability distribution function (pdf) obeys
\begin{equation}\label{eq:pbetha}
    p(\beta_{j}\vert \bm{a}) = p(\mj,\avj \vert \bm{a}) = p(\mj\vert \bm{a})p(\avj).
\end{equation}
For the predicted absolute magnitude prior, we use a normal distribution
\begin{equation}\label{eq:mjk}
    M_{j}^{k}\sim \mathcal{N}(\mathcal{M}_{i}^{k}, \sigma_i^k),
\end{equation}
where the standard deviation $\sigma_i^k$ should be less than half the difference in $k$-band magnitude between adjacent isochrones (see \citetalias{alzate21}, Sec.\,3.2, for details). This tolerance in the isochrone separation is usually a complex function of mass, but, for simplicity, we take it as a constant, averaging $\sigma_i^k$ over all isochrones and all mass values. This does not seem a bad approximation since $\log({\rm luminosity})\propto\log$(age) for pre-MS stars on Hayashi tracks (see Sec.~\ref{sec:sfh}). We adopt $\sigma_i^k$ = 0.01 mag, of the same order of magnitude that the \g photometric errors. A bigger $\sigma_i^k$ results in a smoother SFH with loss of information on the stellar ages.

The prior for parameter $\beta_j$ is then given by
\begin{equation}\label{eq:prior_MG}
p(\beta_{j}\vert \bm{a})\,=\,\sum_{i=1}^{N_{I}}\,\pav\,a_{i}\,\int_{m_{l,i}}^{m_{u,i}}\phi(m)\,\prod_{k=1}^{3} \mathcal{N}(M^{k}_{j}\vert \mathcal{M}^{k}_{i},\sigma_i^k)dm.
\end{equation}

The $\bm{a}$ parameter is directly related to the SFH of the stellar population.
For instance, $a_n=1$ and $a_{i\neq n}=0$ correspond to a population formed with a single burst of star formation. We use the Dirichlet distribution
\begin{equation}\label{eq:dirichlet}
    p(\bm{a})=\frac{\Gamma(\xi N_{I})}{\Gamma(\xi)^{N_{I}}}\prod_{i=1}^{N_{I}}a_{i}^{\xi-1}
\end{equation}
as the prior pdf for $\bm{a}$. This function determines the pdf for random variables which satisfy the conditions $a_{i}$~$\ge$~$0$ and $\sum a_{i}$~=~1, as required in our case. In Eq.~(\ref{eq:dirichlet}), $\Gamma$ and $\xi$ are the Gamma function and the concentration parameter, respectively. In this paper we use $\xi=1$, for which Eq.~(\ref{eq:dirichlet}) returns a uniform pdf, i.e., all $a_{i}$ have the same flat probability distribution.

\subsection{Evolutionary track and isochrone sets}

\begin{figure*}
    \centering
    \includegraphics[width=0.98\textwidth]{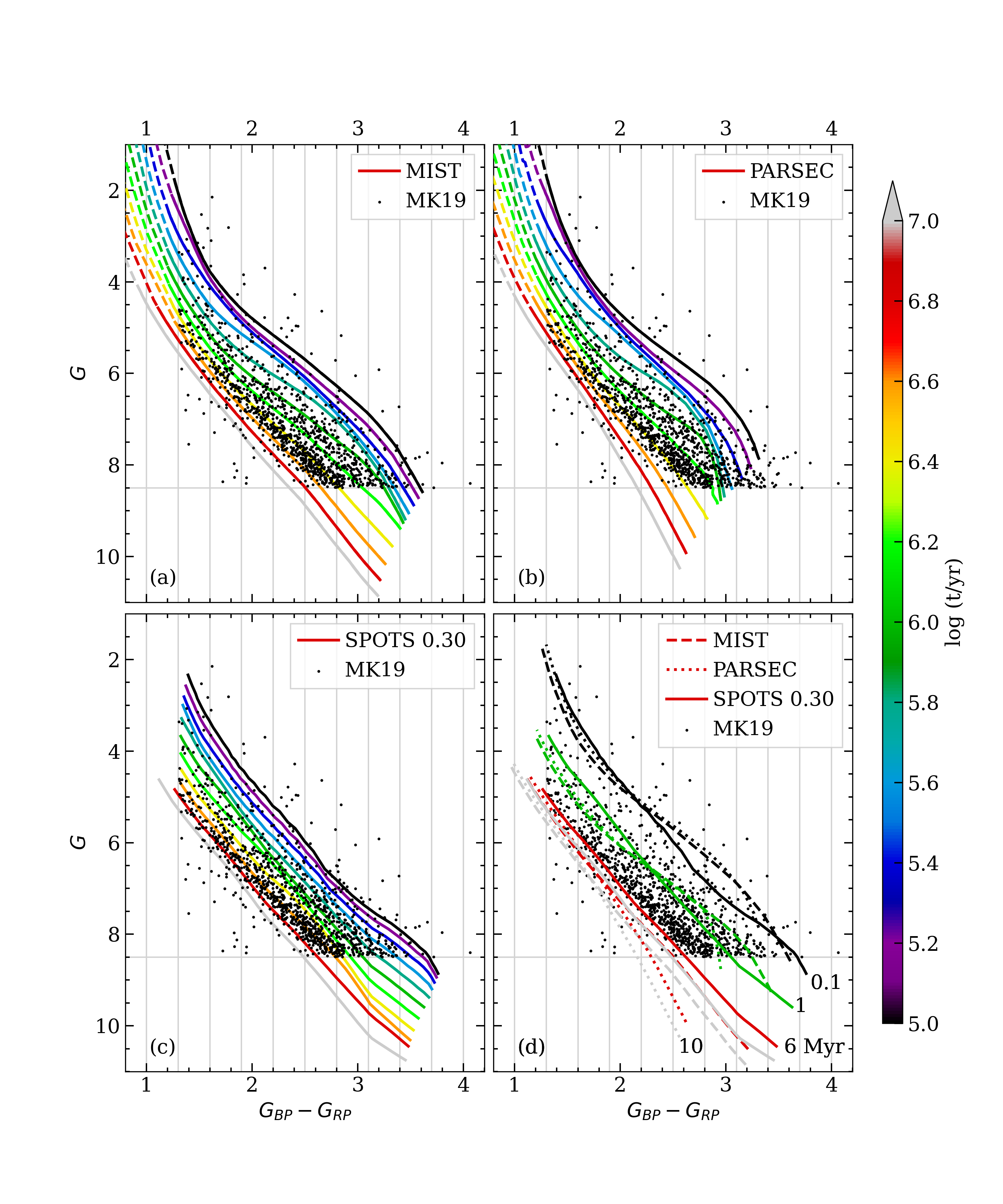}
    \caption{Location of the {\it (a)} MIST, {\it (b)} PARSEC and {\it (c)} SPOTS isochrones in the $G$\,vs.\,$G_{BP}$\,-\,$G_{RP}$ CMD.
    In these panels, 11 isochrones ranging in log\,age\,(yr) from 5 to 7 in steps of 0.2 are plotted as lines, colour coded by age as indicated in the colour bar. 
    The {\it dashed} segment of the MIST and PARSEC isochrones in panels {\it (a)} and {\it (b)} correspond to stellar mass\,>\,1.3\,$\msun$.
    In {\it panel (d)} we compare the 0.1, 1, 6 and 10 Myr isochrones from the three datasets for stellar mass\,$\leq$\,1.3\,$\msun$. In this panel, the age of the isochrones is indicated near their low-mass end. In all panels, the dots correspond to the \citetalias{mck19} stellar sample. A horizontal line is drawn at absolute magnitude $G=8.5$, corresponding to a limiting apparent magnitude $G_{lim}=16.5$ at 400 pc. The vertical lines are drawn to guide the eye when comparing different panels.}
    \label{fig:CMDt}
\end{figure*}

\begin{table*}
\caption{\label{tab:iso}Isochrone grids for $Z=Z_{\odot}$ with solar-scaled abundance ratios.
}
\centering
\begin{tabular}{lcccccccccc}
\hline
 Model & $\log({\rm Age/yr})$ & Step  & $\sigma_{i}$ & N$_{iso}$   & Stellar mass (M$_{\odot}$) &     X    &    Y    &    Z    & $[{\rm Fe}/{\rm H}]$  &  $[{\rm \alpha}/{\rm Fe}]$ \\
\hline                                                                                          
MIST$^a$         & $[5,7]$    &  0.2  & 0.01         &    11       & $[0.1,$\,>\,$20]$     &  0.7155  &  0.2703 &  0.0142 & 0 & 0 \\
\hline                                                                                          
PARSEC 1.1$^b$   & $[5,7]$    &  0.2  & 0.01         &    11       & $[0.1,12]$                &  0.7040  &  0.2790 &  0.0152 & 0 & 0 \\
\hline                                                                                          
SPOTS$^c$        & $[5,7]$    &  0.2  & 0.01         &    11       & $[0.1,1.3]$               &  0.7195  &  0.2676 &  0.0165 & 0 & 0  \\
\hline
\multicolumn{10}{l}{$^a$ \citet{choi2016}, $^b$ \citet{bressan2012}, $^c$ \citet{somers2020,cao21}.}
\end{tabular}
\end{table*}

To interpret the observations in terms of the SFH and the position at birth of each star in the CMD, it is necessary to adopt a specific set of stellar evolutionary tracks (isochrones). Unfortunately, there are considerable differences between different sets of calculations due to uncertainties in the treatment of convection and magnetic activity.
Here we use isochrones derived from the MIST \citep{choi2016}, the PARSEC \citep{bressan2012}, and the SPOTS \citep{somers2020,cao21} sets of evolutionary tracks for solar metallicity, with solar-scaled abundance ratios, and age up to 10 Myr, which include the PMS evolution and provide the \g  $G,\ G_{\rm BP}$ and $G_{\rm RP}$ photometric magnitudes for all the stars along the isochrone (Table~\ref{tab:iso}).
We select 11 isochrones from each set ranging in log age(yr) from 5 to 7 in steps of 0.2.
Fig.~\ref{fig:CMDt} shows the position of these isochrones in the $G$\,vs.\, $G_{BP}$\,-\,$G_{RP}$ CMD together with the \citetalias{mck19} stellar sample.

The MIST and the PARSEC tracks adopt a standard model of stellar evolution, while the SPOTS tracks assume a reduction in the efficiency of convection due to strong
magnetic fields \citep[see also][]{feiden16}. The latter seem to be in better agreement with dynamical mass estimates, mostly from observations of Keplerian rotation in the circumstellar disks of young stars \citep{simon19}. These larger masses tend to make stellar ages larger. The MIST and PARSEC tracks extend to higher mass stars on radiative tracks, where the complications introduced by magnetic fields are likely to be much smaller.

The SPOTS tracks, which incorporate the structural effects of star spots on internal structure, parametrize magnetic effects with a star spot filling factor and spot temperature contrast. In this paper we adopt a filling fraction of $f_{\mathrm{spot}} = 0.3$ and a temperature contrast of $x_{\mathrm{spot}} = 0.8$, values appropriate for Rossby-saturated active stars \citep{cao22}. Reasonable choices of star spot filling factors do not significantly change our estimates---for instance, the choice of $f_{\mathrm{spot}} = 0.5$ does not change the morphology of our SFH, only shifting the mean log age by 0.05 dex toward older ages.

\subsection{Extinction by dust}\label{sec:edust}

\begin{table}
\caption{\label{tab:dust}$\lgg$--$\lga$ groups}
\centering
\begin{tabular}{ccccc}
\hline
 Group & $\lgg$        &   $\lga$         & $\langle \av \rangle$ & c \\
\hline
1      & $\leq 3.6$  &  $\leq 6.04$   & 2.57 & --   \\
2      & (3.6,3.8]   &  (6.04,6.34]   & 2.03 & 0.99 \\
3      & (3.8,4.0]   &  (6.34,6.64]   & 1.76 & 1.14 \\
4      & (4.0,4.2]   &  (6.64,6.94]   & 1.41 & 1.42 \\
5      & $> 4.2$     &  $> 6.94$      & 1.19 & 1.68 \\
\hline
\multicolumn{5}{l}{For group 1 we use $p(\av)=(0.045 * \av + 0.120)/c_{l}$, where}\\
\multicolumn{5}{l}{$c_{l}$ = 0.69 is a normalization constant calculated in the range}\\
\multicolumn{5}{l}{$\av = [0,4.5]$ mag.}\\
\end{tabular}
\end{table}

\begin{figure}
    \centering
    \includegraphics[width=\columnwidth]{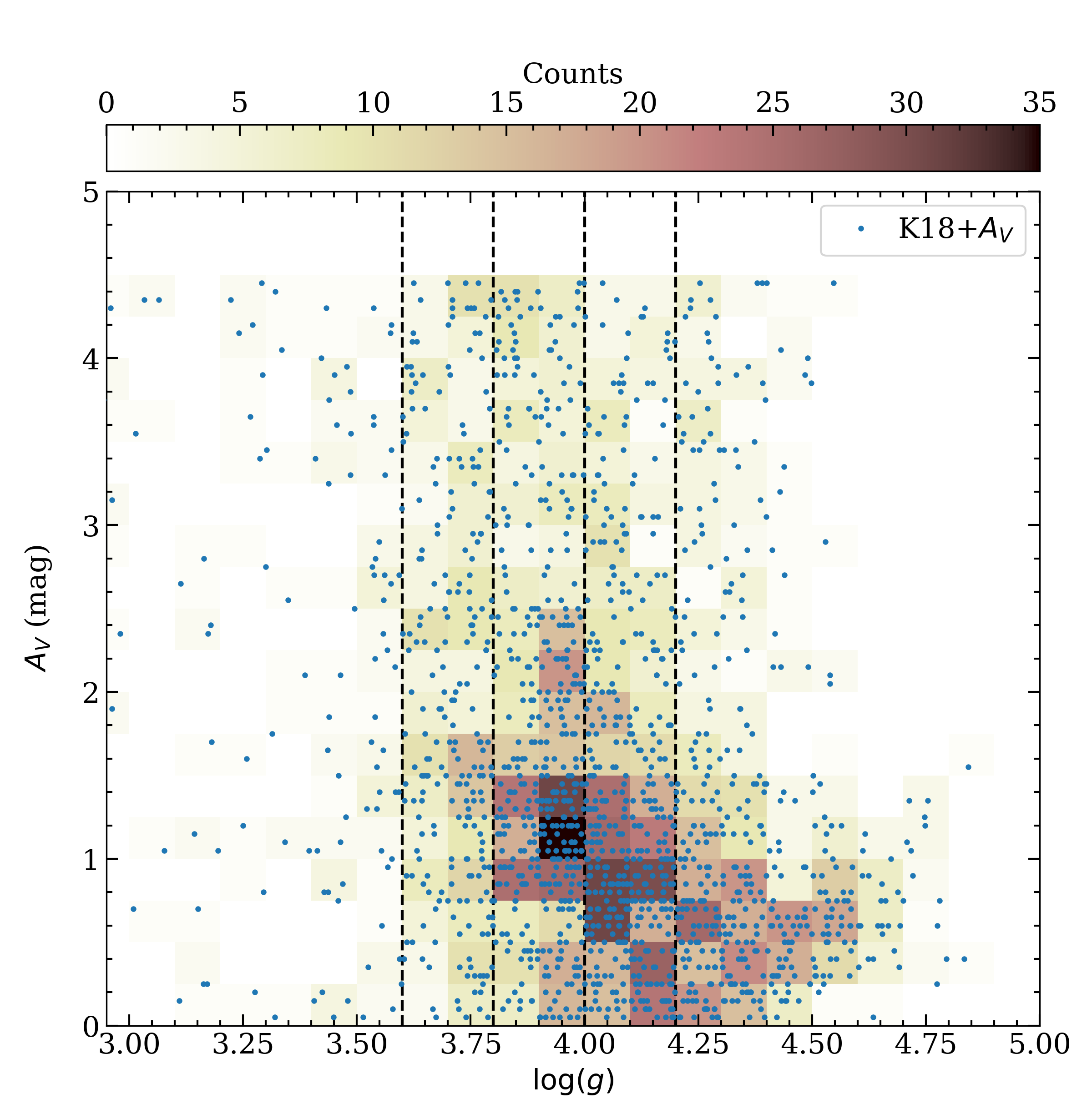}
    \includegraphics[width=\columnwidth]{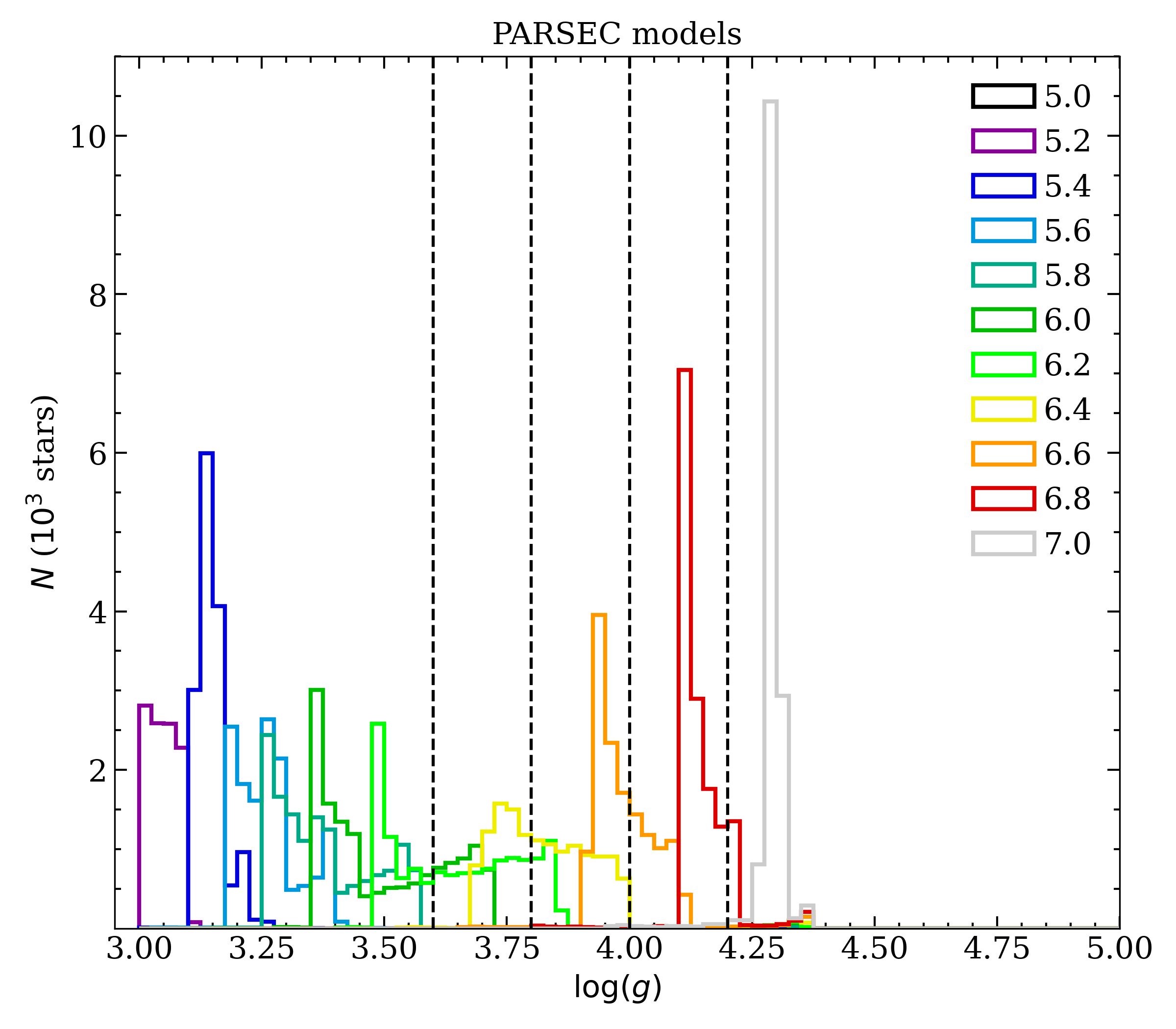}
    \caption{{\it Top panel:} $A_{V}$ vs. $\log(g)$ for the K18 stellar sample.
    {\it Bottom panel:} $\log(g)$ histograms vs. age for a simulated population with constant star formation rate shown colour coded by age at
    $\log({\rm Age/yr})=5, 5.2, 5.4, 5.6, 5.8, 6, 6.2, 6.4, 6.6, 6.8, 7.$
    The {\it dashed} vertical lines are drawn at $\log(g)=3.6, 3.8, 4.0, 4.2$.
    For these simulations we used the PARSEC isochrones and the Kroupa IMF.}
    \label{fig:log.AV.model}
\end{figure}

\begin{figure}
    \centering
    \includegraphics[width=\columnwidth]{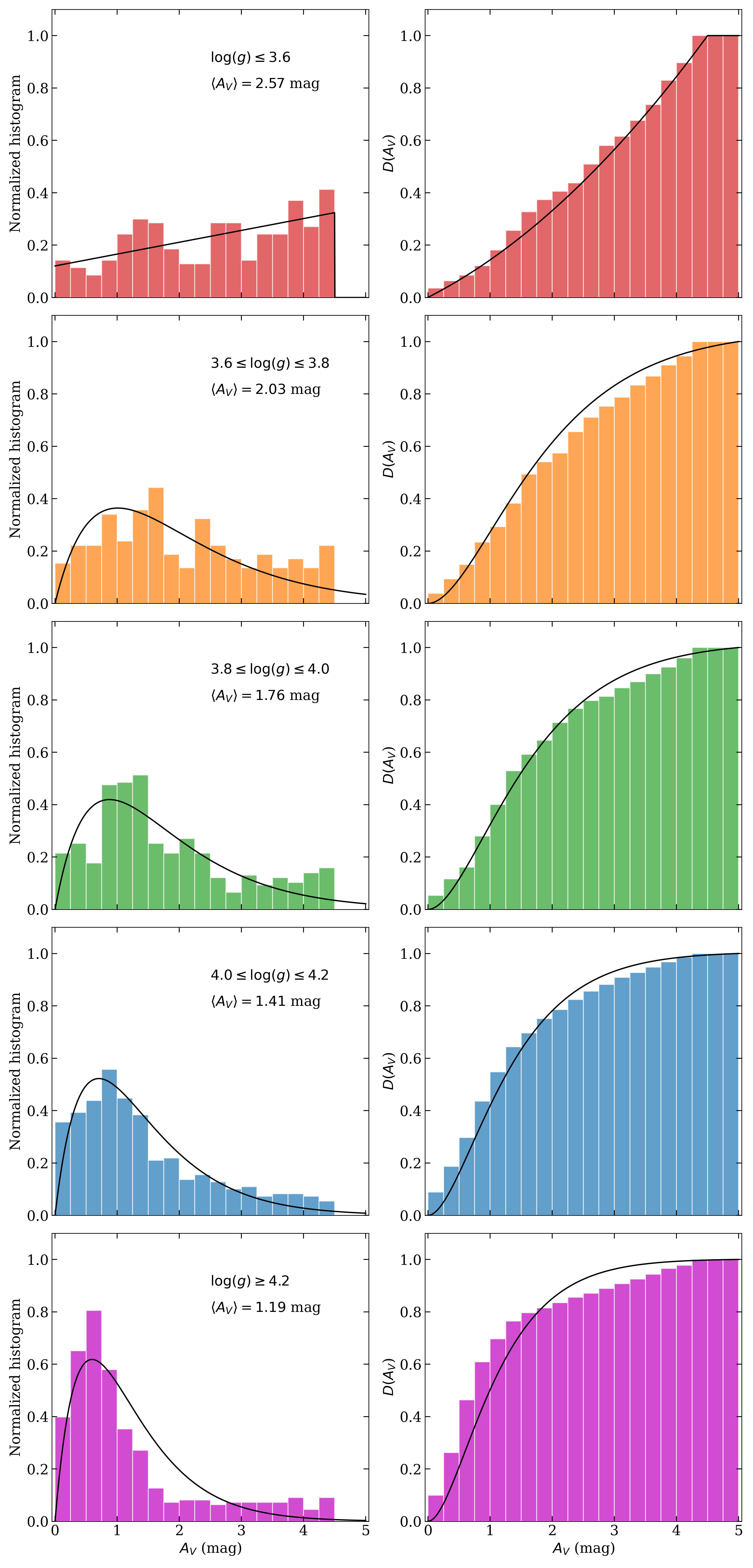}
    \caption{{\it Left panels:} Distribution of $\av$ for the Orion \kss\ stars
    (Sec.~\ref{sec:sample}) grouped according to $\lgg$ as in Table~\ref{tab:dust}. The line corresponds to the fit to Eq.~(\ref{eq:pav}) for groups 2 to 5.
    For group 1 we use $p(\av)=(0.045 * \av + 0.120)/c_{l}$, where $c_{l}$ = 0.69 is a normalization constant calculated in the range $\av = [0,4.5]$ mag.
    {\it Right panels:} Normalized cumulative distribution of $\av$ for the corresponding panel on the {\it left} shown as a histogram and as a continuous line.
    }
    \label{fig:av}
\end{figure}

\begin{figure*}
\centering
\includegraphics[width=0.4\textwidth]{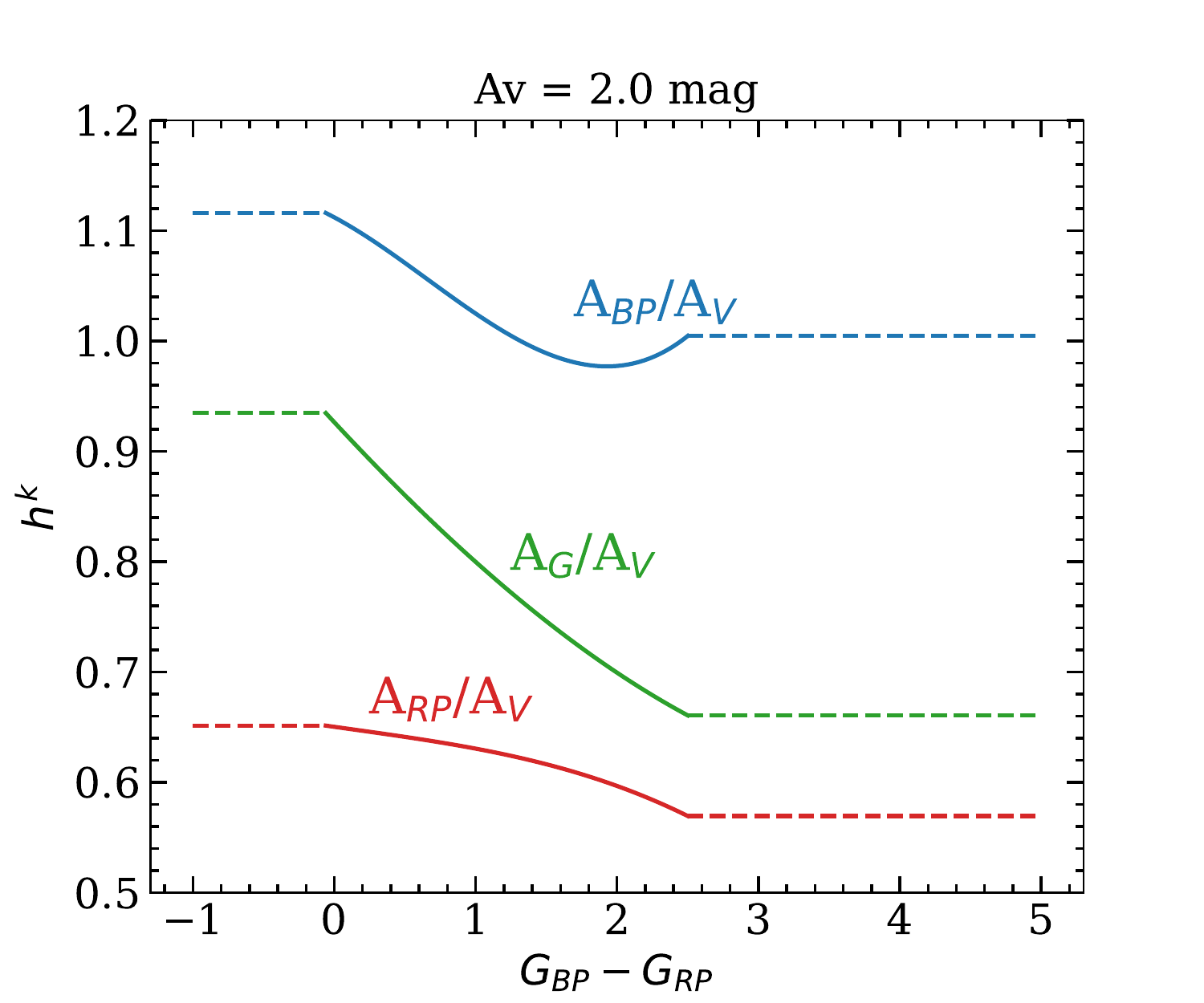}
\includegraphics[width=0.4\textwidth]{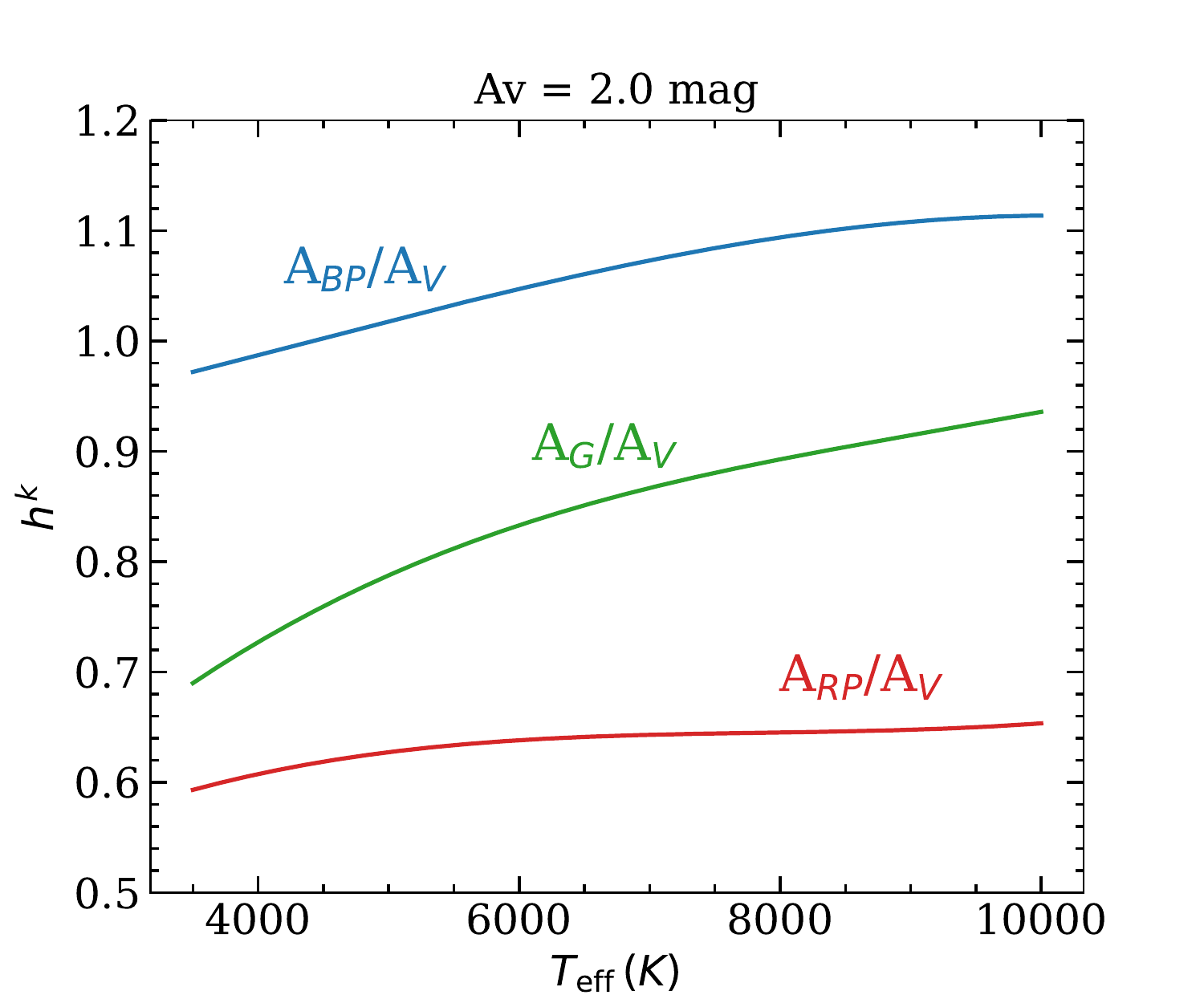}
\caption{\gethree extinction ratios for $\av$~=$~2$ mag vs. intrinsic $G_{BP}-G_{RP}$ colour ({\it left panel}) and effective temperature $3500<T_{\rm eff}/K<10000$ ({\it right panel}).
The {\it solid line} segments show the polynomials provided in 
\url{https://www.cosmos.esa.int/web/gaia/edr3-extinction-law}
for the indicated photometric band. The {\it dashed line} segments are constant extrapolations of the extreme values of the polynomials.
\label{fig:hk}}
\end{figure*}

Since extinction shifts stars to fainter magnitudes and redder colours, {\em roughly} parallel to isochrones in the CMD, corrections are needed for reliable results. However, only a subset of our stellar sample has estimated $\av$ values. The top panel of Fig.~\ref{fig:log.AV.model} shows the distribution of $\av$\,vs.\,$\lgg$ from \kss. In order to use the full sample, we assume that the distribution of extinction for the entire sample is the same as for the stars with measured extinction. 

For the isochrones listed in Table~\ref{tab:iso}, there is a good correlation between the isochrone age and the mean $\lgg$ value of the stars along the isochrone, as shown in the lower panel of Fig.~\ref{fig:log.AV.model} for the PARSEC set. Based on this property, we divide the \kss\ sample in the 5 age/$\lgg$ groups listed in Table~\ref{tab:dust} and build the $\av$ distributions shown in the left hand side panels of Fig.~\ref{fig:av}.
For groups 2 to 5, we fit the $\av$ distribution using the function
\begin{equation}\label{eq:pav}
    p(\av) = c^{2}\cdot\av\cdot e^{-c\cdot\av},
\end{equation}
where $c$ is a constant and $\langle \av \rangle=2/c$. For group 1 we use 
\begin{equation}\label{eq:pav2}
    p(\av) = (0.045 * \av + 0.120)/c_{l},
\end{equation}
where $c_{l}$ = 0.69 is a normalization constant calculated in the range $\av = [0,4.5]$ mag. We adopt the $\pav$ distribution for the $i^{th}$ isochrone according to its age, providing a smooth behaviour of $\av$ for computational purposes even though it is not a perfect fit to the data.

We then compute the extinction in the \g  $G,\ G_{\rm BP}$ and $G_{\rm RP}$ bands using the polynomials provided by the {\it Gaia Collaboration} in \url{https://www.cosmos.esa.int/web/gaia/edr3-extinction-law}. 
Each polynomial is a function of the $G_{\rm BP}-G_{\rm RP}$ colour of the star and its extinction $\av$. Outside the range of validity of the polynomials, $-0.6 < G_{\rm BP}-G_{\rm RP} < 2.5$, we use its value at the closest end. Fig.~\ref{fig:hk} shows the adopted extinction ratios for $\av$\,=\,2.

\subsection{Modeling unresolved binaries}\label{sec:ubin}

\begin{figure}
    \centering
    \includegraphics[width=\columnwidth]{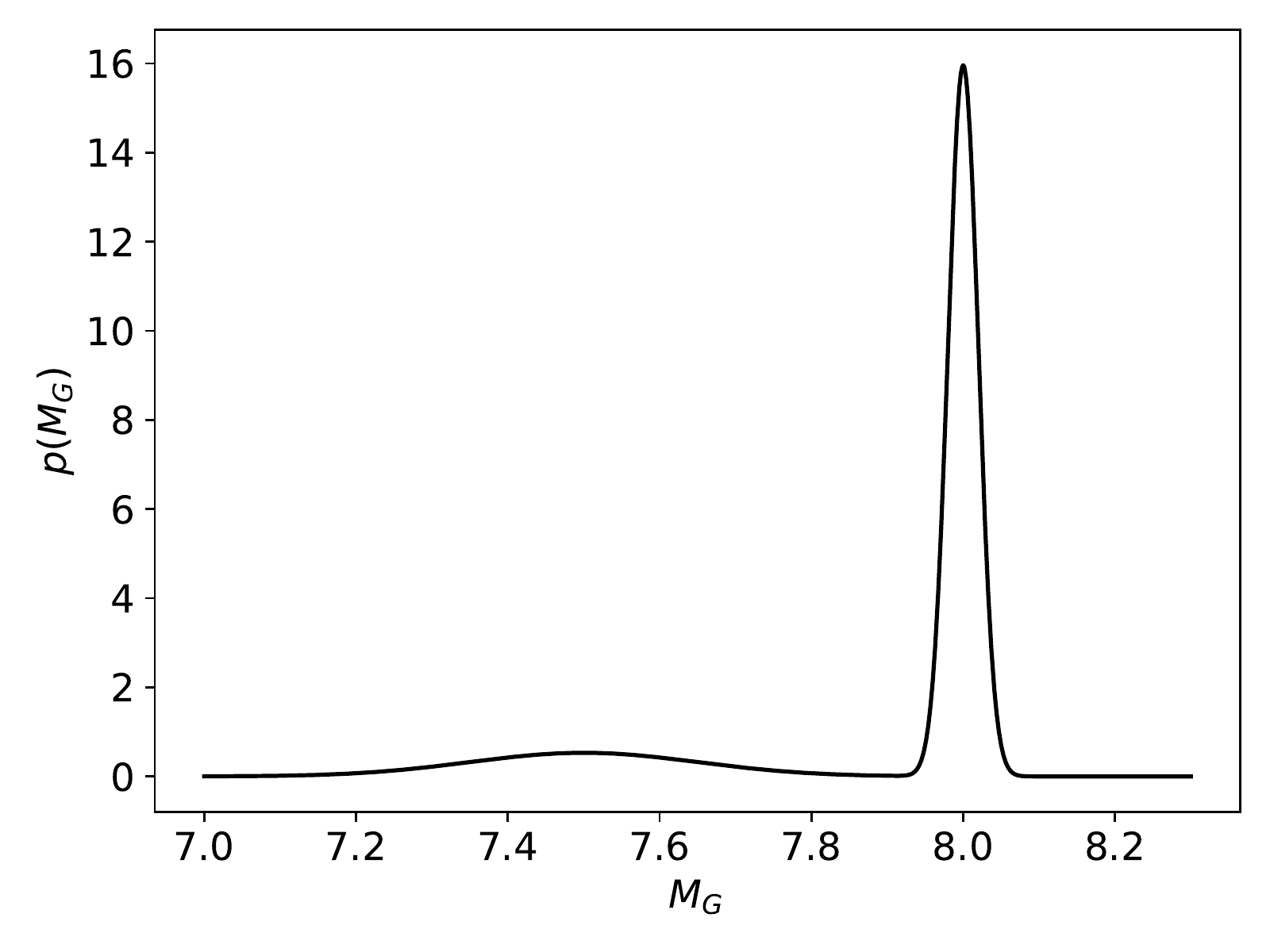}
    \caption{The proper treatment of UBS requires that Eq.~(\ref{eq:mjk}) is replaced by Eq.~(\ref{eq:binary_model}). The figure shows $p(M_G)$ for $M_{i}^{k=1}=8$ mag and $\sigma_{i}=0.02$ mag in Eq.~(\ref{eq:binary_model}). The separation between the two maxima is 0.5 mag.}
    \label{fig:UBS_model}
\end{figure}

The isochrones listed in Table~\ref{tab:iso} are based on stellar evolution models computed for single stars, and hence, will not describe properly the UBS present in the problem sample. UBS are observed in nearby populations \citep{smart20} and most likely they are present in the Orion A sample. UBS are brighter than their component stars considered individually, and they will appear to be younger than their actual age when dated with single star isochrones.
This biases our analysis, introduces systematic errors in our derivation of the SFH and hampers any attempt to constrain the stellar birthline. In the CMD of older nearby star clusters, like the Hyades, a narrow main sequence (MS) is clearly seen. The UBS are easily identified on the binary sequence that is parallel to the MS \citep{smart20}. In this case, it is possible to estimate the number fraction of UBS, and the bias in age or metallicity can be treated.
Most of the Orion A stars are in the pre-MS and show a wide distribution in luminosity (age) at any given colour, making it challenging to disentangle UBS from single stars. To address this issue, we use a simulated mock catalogue of the Orion A stellar population (Appendix~\ref{app:basis}) to quantify the biases introduced in our results by ignoring UBS. In the simulation we control all the stellar inputs, including the binary population properties.

The exercise of recovering the SFH from the mock catalogue helped us understand the biases in the inferred stellar ages. We must replace Eq.~(\ref{eq:mjk}) by
\begin{equation}\label{eq:binary_model}
    M_{j}^{k}\sim 0.8\cdot\mathcal{N}\left(\mathcal{M}_{i}^{k}, \sigma_i^k\right) + 0.2\cdot\mathcal{N}\left([\mathcal{M}_{i}^{k}-0.5], 0.15\right),
\end{equation}
to model properly UBS (cf. Fig~.\ref{fig:UBS_model}). We add a second normal distribution to extend the reach of the isochrones towards brighter stars (ignoring triple and higher order systems). The mean ($\mathcal{M}_{i}^{k}-0.5$) and the standard deviation ($0.15$ mag) of the additional component were selected based on careful examination of the CMDs of the NGC 2232 and the Hyades open clusters \citep{smart20}. In Eq.~(\ref{eq:binary_model}) we assume that 80\% of the stars in the CMD are single (including resolved binaries) and 20\% are UBS. Each group of 10 stars in a narrow $G_{BP}-G_{RP}$ colour range will contain 2 UBS, for a total of 12 stars in the group. We note that these are not the total number of UBS but merely an approximation of the effect on the CMD. We have tested this assumption using the mock catalogue (Appendix A), which incorporates a detailed model of binary statistics derived from studies of field stars.
We find that Eq.~(\ref{eq:binary_model}) helps reducing spurious contributions to the SFH at young ages.

\subsection{Marginalized posterior pdf}\label{sec:marg_post}

The usefulness of our Bayesian hierarchical model in this study rests on its ability to infer population parameters using simultaneously observations of a large number of stars. We do not infer the age of individual stars, but from the posterior pdf we infer the SFH of the whole population, from which star ages can be estimated. The hierarchy of our statistical model establishes that the population parameters $\{a_i\}$ and the priors ($\phi$, $p(\av)$, isochrones) determine the distribution of $\mj$ and $\avj$, from which we derive the observables through Eq.~(\ref{eq:pgj}). In this scheme, the individual parameters are the connection between the observables and the population parameters, where only the latter matter for the computation of the SFH. In other words, the posterior pdf of each individual parameter is needed but not relevant for the final result. Marginalizing (integrating) the posterior pdf respect to the $\bm{\beta}=(\mj,\avj)$ parameters, we obtain the distribution of {\bf a} given the data {\bf d},
\begin{equation}\label{eq:marg_post}
    P(\bm{a}\vert \bm{d}) = \int P(\bm{a}, \bm{\beta} \vert \bm{d},\phi) d\bm{\beta}.
\end{equation}
For a complete sample, the completeness function $S(d_{j})$ in Eq.~(\ref{eq:bayes}) fulfills $S(d_{j})\approx 1$ and no significant number of stars is lost to a given brightness limit. The integration limits in Eq.~(\ref{eq:marg_post}) are then
\begin{equation}\label{eq:comp_int}
\avj=[0,\infty]\ \ \ {\rm and}\ \ \ \mj=\left[-\infty,\infty\right].
\end{equation}
If the sample is complete up to apparent magnitude $G=G_{\rm lim}$, we compute the integral in 
Eq.~(\ref{eq:marg_post}) in the interval
\begin{equation}\label{eq:mag_cut}
\mj=\left[-\infty,G_{\rm lim}^{k}-f(\varpi_{j})\right],
\end{equation}
where $f(\varpi_{j})$ is the distance modulus defined in Eq.~(\ref{eq:dm}). This brightness limit makes the posterior pdf go to zero for $G_{j}>G_{\rm lim}$.
Thus, the statistical model predicts a null number of stars bellow the magnitude cut-off. By construction, the normalization of the posterior includes this magnitude cut-off, the stellar counts predicted by the posterior are re-normalized and will follow the IMF distribution for stars within the allowed brightness range.

At this point, Bayes's theorem has been fully implemented within our framework.
We can now proceed to infer the marginal posterior pdf of $a_{i}$ performing a Markov chain Monte Carlo (MCMC) process using, for instance, the Stan platform (\url{https://mc-stan.org}).
We use the median of the marginal posterior pdf ($a50_i$) as the estimator of the fraction of stars associated to the $i^{th}$ isochrone.
The 10 and 90\% percentiles ($a10_i$ and $a90_i$, respectively) enclose the 80\% credibility region of the inference.

\subsection{Number and mass in stars per isochrone}\label{sec:nummass}

We use the median of the marginalized distribution of $a_{i}$ as an unbiased estimator of the number of stars ($N_{*}^i$) assigned to the $i^{th}$ isochrone by our statistical model. 
The credibility interval of $N_{*}^{i}$ is derived from the 10 and 90\% percentiles of this distribution, denoted
$a_{10,i}$ and $a_{90,i}$.
The total mass of these stars is the product of $N_{*}^i$ times the average stellar mass,
\begin{equation}\label{eq:mmean}
    \langle m\rangle=\frac{\int_{m_l}^{m_u} m \phi(m) dm} {\int_{m_l}^{m_u} \phi(m) dm},
\end{equation}
where $\phi(m)$ is the IMF. Throughout this paper we use the \citet{Kroupa2001} IMF, integrating $\phi(m)$ from $m_{l}$\,=\,0.1 to $m_{u}$\,=\,100\,$\msun$.
In this case $\langle m\rangle$\,=\,0.6377\,$\msun$. The SFR $\Psi(t_i)$ then follows from the ratio
\begin{equation}\label{eq:sfr}
    \Psi(t_i) =
    \frac{\langle m\rangle\,N_{*}^i}{\Delta t_i} =
    \frac{\langle m\rangle\,N_{*}^i}{2.30258\,t_i\,\Delta({\rm log}_{10}\,t_i)}.
\end{equation}
where $\Delta t_i$\,=\,$(t_{i+1}-t_{i-1})/2$ and the rightmost term allows for an isochrone grid spaced logarithmically in time. We note that the integration down to $m_{l}=0.1\,\msun$ in Eq.~(\ref{eq:mmean}) is meaningful only when the sample is complete to this mass limit.\footnote{The integrals in Eqs.~(\ref{eq:mmean}, \ref{eq:mmeanl}) are insensitive to $m_{u}$ as long as $m_{u}$\,>>\,1\,$\msun$.} If the sample is complete only to magnitude $G_{\rm lim}$, the average $\langle m\rangle$ is defined as
\begin{equation}\label{eq:mmeanl}
    \langle m\rangle_i=\frac{\int_{m_i}^{m_u} m \phi(m) dm} {\int_{m_i}^{m_u} \phi(m) dm},
\end{equation}
\noindent where $m_i$\,=\,$m(G_{\rm lim},t_i)$ is the stellar mass at $G$\,=\,$G_{\rm lim}$ for the $i^{th}$ isochrone and we have assumed that $\av$\,=\,0 for all the stars. $\langle m\rangle_i$ is then used in Eq.\,(\ref{eq:sfr}) to infer the SFH for magnitude-limited samples corrected for incompleteness.

\subsection{Corrections for effects of extinction}\label{sec:corr_ext}

The correction of the stellar counts when we deal with magnitude limited samples (using Eq.~\ref{eq:mag_cut}) works properly if $\av$\,=\,0 for all the stars. This is not true when $\av$\,$\neq$\,0 for two reasons. First, the reddening vectors are not equally parallel to all isochrones, especially for hotter stars. Second, the effect of extinction causes some stars to become fainter than $G_{\rm lim}$. This loss of information results in biased inferences of the stellar fractions ($a_i$). It would be expected that if $p(\av)$ was known, the mean number of attenuated stars could be estimated. However, the wide magnitude range occupied by pre-MS stars in the CMD, the dependence of $A^k/\av$ on the $G_{BP}-G_{RP}$ colour and the randomness of $\av$, make this estimation a non trivial task.
For this reason, in Sec.~\ref{app:incompcorr} we derive an empirical-theoretical incompleteness correction as follows. We simulate stellar populations of $2\times 10^{5}$ single stars, formed at various rates during 10 Myr, resulting on $\mathcal{N}_{{\rm c}}(t_i)$ stars per age bin, where $t_i$ is the $i^{th}$ isochrone age. Each star is attenuated by an amount $\av$ sampled from the $p(\av)$ pdf. The corrected number of stars in the $i^{th}$ bin follows from
\begin{equation}\label{eq:ai_cor}
  N50'_i = \epsilon_i\cdot N50_i,
\end{equation}
\noindent where the incompleteness correction $\epsilon$ is defined in Eq.~(\ref{eq:incomp}), listed in Table\,\ref{tab:imass} and plotted in Fig.\,\ref{fig:incompleteness}. Throughout this paper we assume, based on Fig.~\ref{fig:hist_g}, that
\begin{equation}\label{eq:g_lim}
G_{\rm lim} = 16.5\ {\rm mag}.
\end{equation}

In summary, if $\av$\,>\,0 we use Eqs.\,(\ref{eq:marg_post}, \ref{eq:comp_int}, and \ref{eq:ai_cor}) to calculate $a_i$ and 
Eqs.\,(\ref{eq:sfr} and \ref{eq:mmeanl}) to derive the SFH corrected for incompleteness and interstellar reddening.

\section{Star Formation in Orion A}\label{sec:results}

Having shown in Appendix~\ref{app:basis} that we can accurately retrieve the SFH of the mock catalogue with our methods, we now turn to inferring the SFH of the Orion A population. The incompleteness correction applied to the stellar counts inferred from magnitude limited samples is derived in Sec.~\ref{app:incompcorr}.

\subsection{Assessing our treatment of extinction}

\begin{figure*}
\begin{center}
    \includegraphics[width=\textwidth]{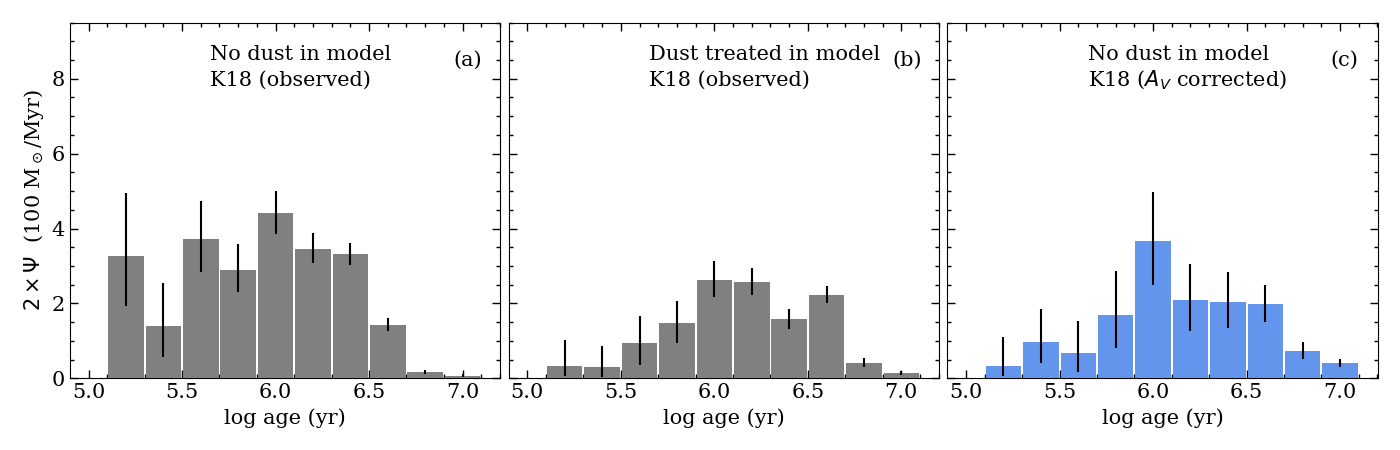}
    \caption{SFH inferred from the 620 stars in the \citetalias{kounkel18} sample with measured $\av$ using the MIST isochrones.
    The stellar sample was corrected by extinction before performing the inference shown in panel {\it (c)}.
    UBS were ignored for this test.
    The values of $\Psi$ have been multiplied by 2 to plot them in the same scale of Fig.~\ref{fig:sfh_results}.
    The results in panel {\it (b)} have been corrected for sample incompleteness (Sec.~\ref{app:incompcorr}).
    The median values and error bars (credibility intervals defined in Sec.\,\ref{sec:nummass}) are listed in columns \ref{fig:sfh_results_b}\,{\it (a,b,c)} of Table\,\ref{tab:resFig6}.
    \label{fig:sfh_results_b}}
\end{center}
\end{figure*}

\begin{figure}
\begin{center}
    \includegraphics[width=0.9\columnwidth]{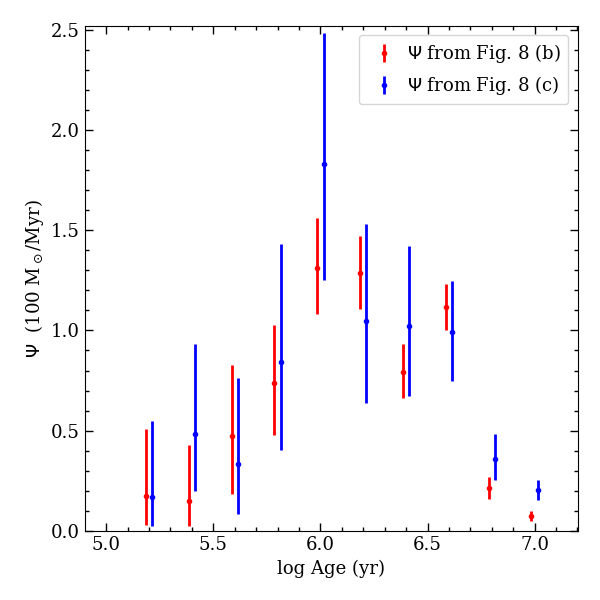}
    \caption{Comparison between the SFHs from panels {\it (b)} and {\it (c)} of Fig.\,\ref{fig:sfh_results_b}.
    The median values and error bars (credibility intervals defined in Sec.\,\ref{sec:nummass}) are listed in columns \ref{fig:sfh_results_b}\,{\it (b)} and \ref{fig:sfh_results_b}\,{\it (c)} of Table\,\ref{tab:resFig6}.
    \label{fig:sfh_results_bc}}
\end{center}
\end{figure}

To assess the performance of the statistical treatment of extinction introduced in Sec.~\ref{sec:edust}, we compare the SFH inferred from stars with  
{\it (i)} photometry corrected by extinction and
{\it (ii)} uncorrected data, modelling extinction as indicated in Sec.~\ref{sec:edust}.
This test requires a sample of stars for which the effective temperature and $A_{V}$ have been determined, such as the \kss\ sub-sample used in Sec.~\ref{sec:edust}.
Panel {\it (a)} of Fig.\,\ref{fig:sfh_results_b} shows the inferred SFH from \kss\ as observed,
with no extinction correction applied to the data nor introduced in our statistical model.
Panel {\it (b)} shows the SFH of \kss\ when we use Eq.~(\ref{eq:pav}) to introduce the effects of extinction statistically in our model.
Panel {\it (c)} shows the inferred SFH, this time correcting the data by extinction using $\av$ from \citetalias{kounkel18}.
Ignoring extinction yields large number of stars with $t$(yr)\,<\,6 dex: extinction moves stars towards the region occupied by young objects (protostars) in the CMD, and heavily reddened stars are then assigned young ages. The inferred SFR decreases dramatically for $t$(yr)\,<\,5.5 -- 6 dex when either the data are corrected for reddening ({\it panel c}) or extinction is included in our statistical model ({\it panel b)}.

The credibility intervals (plotted as error bars) in Fig.\,\ref{fig:sfh_results_b}{\it (c)} are larger than in Fig.\,\ref{fig:sfh_results_b}{\it (b)} due to the uncertainty on $\av$.
The SFHs in Figs.\,\ref{fig:sfh_results_b} {\it (b,c)}, replotted in Fig.\,\ref{fig:sfh_results_bc}, deserve special attention because from their comparison we can assess our statistical model. 
We note in Fig.\,\ref{fig:sfh_results_bc} that the allowed range of the SFH for the two solutions overlap for most age bins. 
Only at log age(yr)\,=\,7 the credibility
bars do not overlap. At log age(yr)\,=\,6.8 the bars touch at their extremes.
Since, by definition, the {\it true} $\Psi(t_i)$ lies within the credibility interval with 80\% probability, we conclude that our statistical correction of the extinction yields a SFH which is consistent with the SFH inferred from data previously corrected by extinction at least up to log age(yr)\,=\,6.8.

\subsection{Results for Orion A}

\begin{figure*}
\begin{center}
    \includegraphics[width=0.9\textwidth]{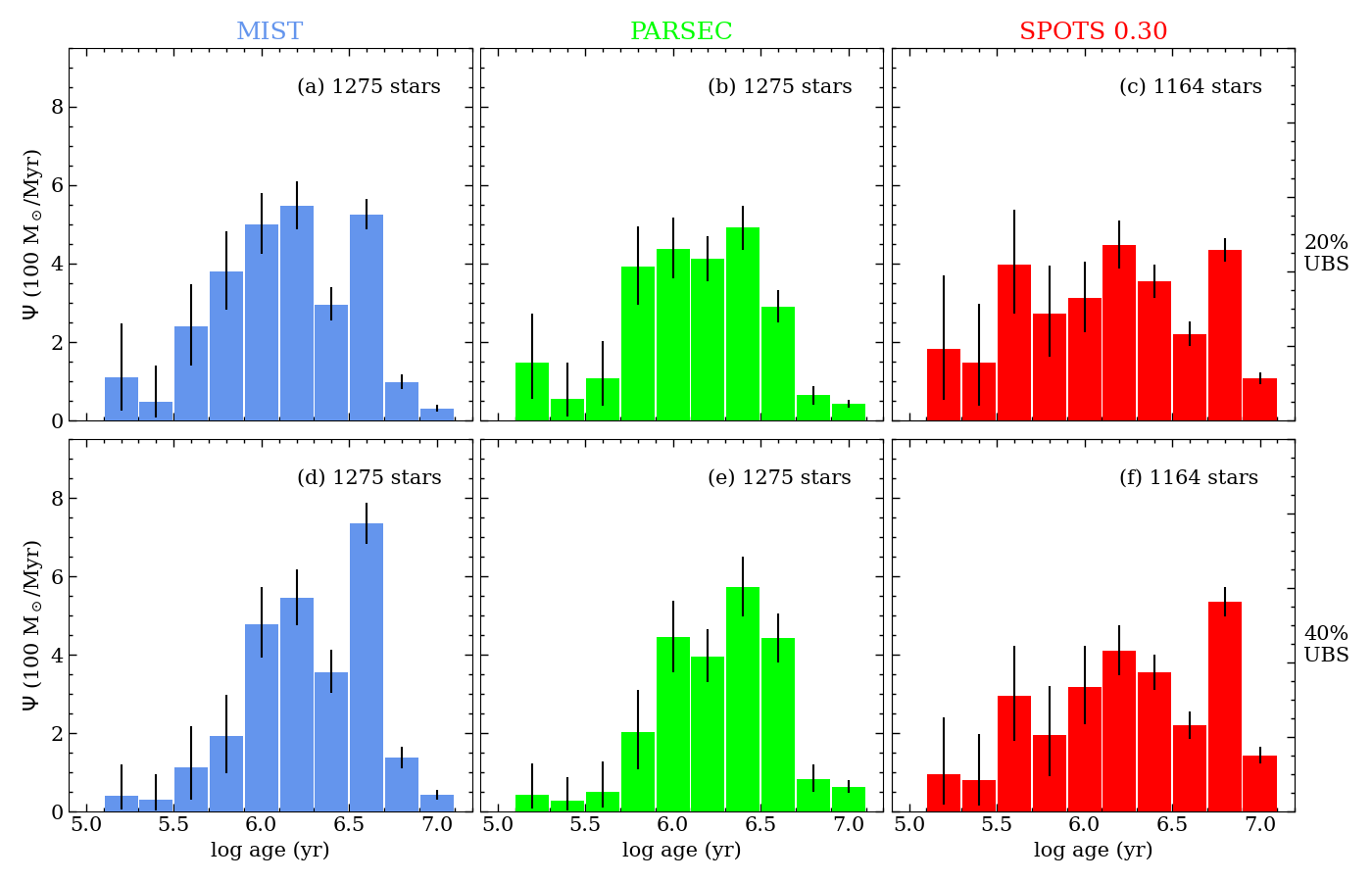}
    \caption{SFH inferred from the \citetalias{mck19} sample with $G<=16.5$ mag (1275 stars) using the MIST and the PARSEC isochrones ({\it blue} and {\it green histograms} respectively),
    and from the MK19 sample with $G<=16.5$ mag and $G_{BP}-G_{RP}>=1.3$ mag (1164 stars) using the SPOTS 0.30 isochrones (red histograms).
    Panels (a,b,c) show the results when we model {20\%} of UBS, and panels (d,e,f) when the UBS fraction is {40\%}. The SFHs for SPOTS were inferred centering the second normal distribution in Eq. (12) in $M_i^k=0.65$ mag instead of $M_i^k=0.5$ mag.
    All results have been corrected for sample incompleteness (Sec.\,\ref{app:incompcorr}).
    The median values and error bars (credibility intervals) are listed in columns \ref{fig:sfh_results}\,{\it (a,b,c)} of Table\,\ref{tab:resFig6}.
    \label{fig:sfh_results}}
\end{center}
\end{figure*}

\begin{figure*}
\begin{center}
    \includegraphics[width=\textwidth]{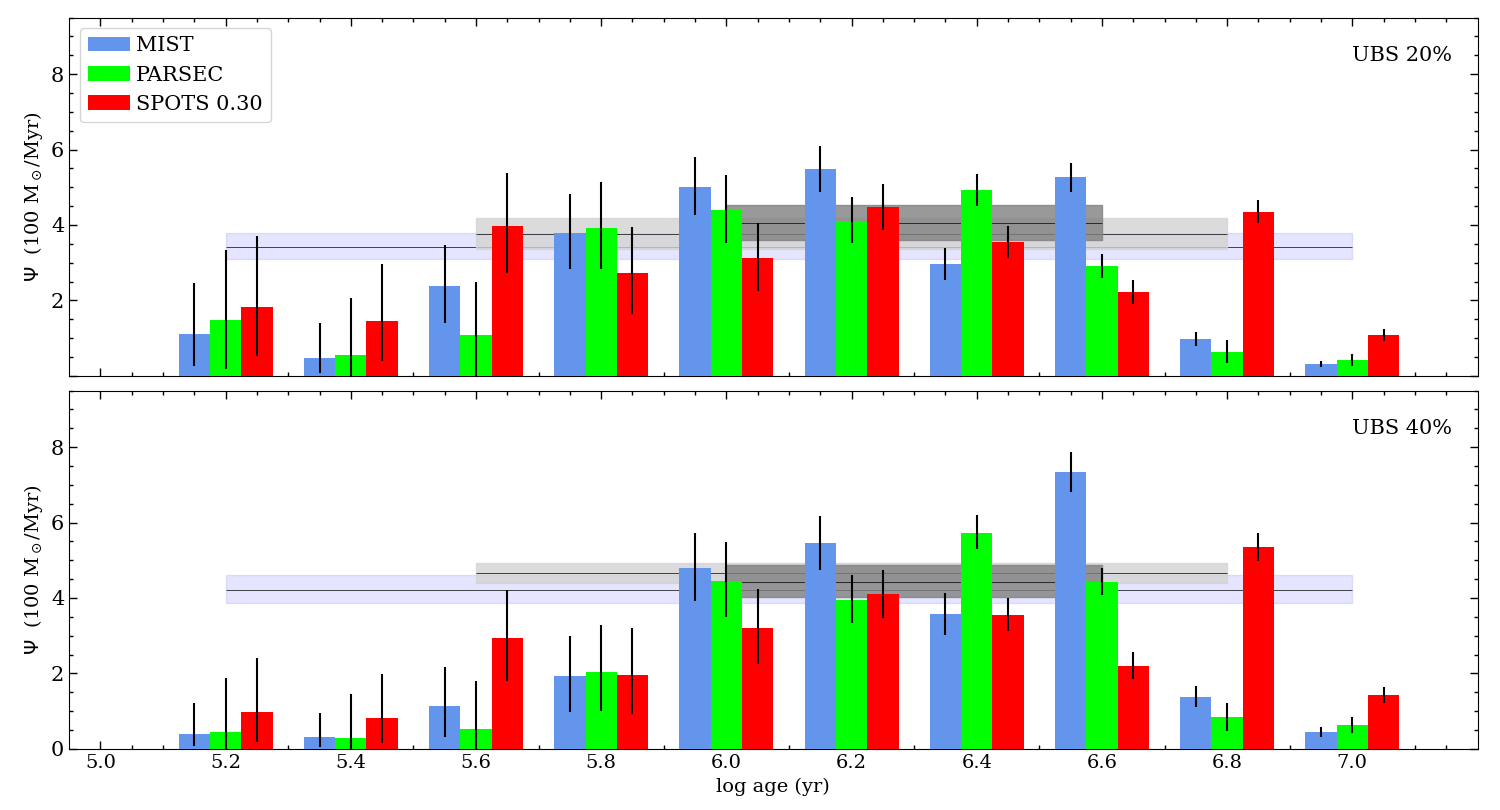}
    \caption{The SFHs from Fig.\,\ref{fig:sfh_results} are plotted next to each other, shifted in log age for easy comparison.
    \textbf{The {\it top panel} corresponds to 20\% UBS and the {\it bottom panel}  to 40\% UBS.}
    The {\it mass-weighted} average SFR $\langle \Psi_{mw} \rangle$ for 3 different intervals of log age for the 3 
    solutions combined are plotted as {\it gray} colour horizontal bands (see Table\,\ref{tab:aveFig6} for details).
    To a first approximation, $\langle \Psi_{mw} \rangle\,\approx\,380\,\pm\,70\,{\rm M}_\odot\,{\rm Myr}^{-1}$ \textbf{for 20\% UBS}.
    This is a lower limit for $\Psi$ since heavily reddened stars not visible in the Gaia bands have not been included in its estimate.
    \label{fig:sfh_summary}}
\end{center}
\end{figure*}

\begin{table*}
  \begin{center}
  \renewcommand{\arraystretch}{1.48} 
  \caption{\label{tab:resFig6}Orion A star formation history from model inferences (Figs.~\ref{fig:sfh_results_b} and \ref{fig:sfh_results}).}
  \begin{tabular}{crcccccccccccc}
  \hline
       & Sample:         & & \citetalias{kounkel18}$^a$ & \citetalias{kounkel18}$^a$  & \citetalias{kounkel18}$^b$  &  &  \citetalias{mck19}$^c$ & \citetalias{mck19}$^c$ & \citetalias{mck19}$^d$  &  &  \citetalias{mck19}$^c$ & \citetalias{mck19}$^c$ & \citetalias{mck19}$^d$ \\
       & Isochrones:      & &  MIST             &   MIST                   &  MIST                  &  &  MIST                       &   PARSEC                  &   SPOTS                   &  &  MIST                       &   PARSEC                  &   SPOTS\\
       & Fig.\,(panel):  & &  \ref{fig:sfh_results_b}\,{\it (a)} & \ref{fig:sfh_results_b}\,{\it (b)} & \ref{fig:sfh_results_b}\,{\it (c)}  &  & \ref{fig:sfh_results}\,{\it (a)} & \ref{fig:sfh_results}\,{\it (b)} &  \ref{fig:sfh_results}\,{\it (c)}  &  & \ref{fig:sfh_results}\,{\it (d)} & \ref{fig:sfh_results}\,{\it (e)} &  \ref{fig:sfh_results}\,{\it (f)} \\
\cline{1-2} \cline{4-14}
log age &  age (Myr) &  &  \mlc{11}{c}{$\Psi$ (100 M$_\odot$/Myr)} \\
\cline{1-2} \cline{4-14}
   5.2 &  0.158 &  &   $1.64_{-0.67}^{+0.84}$  &  $0.17_{-0.14}^{+0.34}$  & $0.17_{-0.14}^{+0.38}$  &  &   $1.09_{-0.83}^{+1.34}$   &   $1.46_{-0.92}^{+1.23}$   &   $1.65_{-1.17}^{+1.67}$  &  &   $0.40_{0.34-}^{+0.81}$   &   $0.44_{-0.37}^{+0.80}$   &   $0.97_{-0.78}^{+1.43}$  \\
   5.4 &  0.251 &  &   $0.70_{-0.41}^{+0.57}$  &  $0.15_{-0.13}^{+0.28}$  & $0.48_{-0.28}^{+0.45}$  &  &   $0.48_{-0.40}^{+0.91}$   &   $0.54_{-0.45}^{+0.90}$   &   $1.31_{-0.96}^{+1.35}$  &  &   $0.30_{-0.26}^{+0.65}$   &   $0.29_{-0.25}^{+0.60}$   &   $0.81_{-0.66}^{+1.17}$  \\
   5.6 &  0.398 &  &   $1.87_{-0.45}^{+0.51}$  &  $0.47_{-0.29}^{+0.36}$  & $0.33_{-0.25}^{+0.43}$  &  &   $2.35_{-0.97}^{+1.07}$   &   $1.07_{-0.70}^{+0.92}$   &   $3.57_{-1.13}^{+1.26}$  &  &   $1.13_{0.81-}^{+1.05}$   &   $0.52_{-0.42}^{+0.77}$   &   $2.95_{-1.14}^{+1.27}$  \\
   5.8 &  0.631 &  &   $1.45_{-0.30}^{+0.34}$  &  $0.74_{-0.26}^{+0.29}$  & $0.84_{-0.44}^{+0.59}$  &  &   $3.73_{-0.95}^{+1.02}$   &   $3.85_{-0.95}^{+1.01}$   &   $2.44_{-0.97}^{+1.11}$  &  &   $1.93_{0.96-}^{+1.06}$   &   $2.05_{-0.96}^{+1.07}$   &   $1.97_{-1.05}^{+1.25}$  \\
   6.0 &  1.000 &  &   $2.21_{-0.28}^{+0.30}$  &  $1.31_{-0.23}^{+0.25}$  & $1.83_{-0.58}^{+0.65}$  &  &   $4.91_{-0.73}^{+0.79}$   &   $4.31_{-0.75}^{+0.79}$   &   $2.80_{-0.77}^{+0.85}$  &  &   $4.79_{-0.87}^{+0.93}$   &   $4.45_{-0.90}^{+0.94}$   &   $3.20_{-0.96}^{+1.04}$  \\
   6.2 &  1.585 &  &   $1.73_{-0.20}^{+0.21}$  &  $1.29_{-0.18}^{+0.19}$  & $1.05_{-0.41}^{+0.48}$  &  &   $5.38_{-0.58}^{+0.60}$   &   $4.04_{-0.55}^{+0.57}$   &   $4.01_{-0.54}^{+0.56}$  &  &   $5.44_{-0.70}^{+0.73}$   &   $3.96_{-0.66}^{+0.69}$   &   $4.10_{-0.62}^{+0.65}$  \\
   6.4 &  2.512 &  &   $1.66_{-0.15}^{+0.15}$  &  $0.79_{-0.13}^{+0.14}$  & $1.02_{-0.35}^{+0.40}$  &  &   $2.91_{-0.40}^{+0.42}$   &   $4.83_{-0.55}^{+0.55}$   &   $3.17_{-0.36}^{+0.38}$  &  &   $3.57_{-0.54}^{+0.58}$   &   $5.73_{-0.76}^{+0.77}$   &   $3.55_{-0.43}^{+0.45}$  \\
   6.6 &  3.981 &  &   $0.72_{-0.08}^{+0.09}$  &  $1.11_{-0.11}^{+0.12}$  & $0.99_{-0.24}^{+0.25}$  &  &   $5.17_{-0.37}^{+0.37}$   &   $2.86_{-0.41}^{+0.42}$   &   $1.99_{-0.27}^{+0.29}$  &  &   $7.34_{-0.53}^{+0.54}$   &   $4.43_{-0.62}^{+0.62}$   &   $2.21_{-0.34}^{+0.36}$  \\
   6.8 &  6.310 &  &   $0.09_{-0.02}^{+0.03}$  &  $0.21_{-0.05}^{+0.06}$  & $0.36_{-0.10}^{+0.12}$  &  &   $0.96_{-0.18}^{+0.19}$   &   $0.63_{-0.23}^{+0.24}$   &   $3.90_{-0.27}^{+0.28}$  &  &   $1.38_{-0.28}^{+0.29}$   &   $0.84_{-0.35}^{+0.36}$   &   $5.35_{-0.37}^{+0.38}$  \\
   7.0 & 10.000 &  &   $0.04_{-0.01}^{+0.02}$  &  $0.07_{-0.02}^{+0.03}$  & $0.20_{-0.05}^{+0.05}$  &  &   $0.30_{-0.08}^{+0.09}$   &   $0.41_{-0.09}^{+0.11}$   &   $0.96_{-0.14}^{+0.15}$  &  &   $0.44_{-0.12}^{+0.14}$   &   $0.63_{-0.15}^{+0.17}$   &   $1.43_{-0.21}^{+0.22}$  \\
\hline
\multicolumn{11}{l}{$^a$620 stars with measured $\av$ in the \citetalias{kounkel18} sample. $^b$Stellar sample corrected by extinction before inferring $\Psi(t)$.} \\
\multicolumn{11}{l}{$^c$1275 stars from the \citetalias{mck19} sample. $^d$1164 stars with $G_{BP}-G_{RP}$\,$\geq$\,1.3 mag. For all samples, $G_{\rm lim}$\,=\,16.5 mag.}  \\
\end{tabular}
\end{center}
\end{table*}
\begin{table}
  \begin{center}
  \renewcommand{\arraystretch}{1.48} 
  \caption{\label{tab:aveFig6}Orion A mass-weighted SFH from model inferences in Fig.~\ref{fig:sfh_summary}}
  \begin{tabular}{cccccc}
  \hline
  \multirow{2}{*}{UBS} & Range       &  \mlc{4}{c}{$\langle\,\Psi_{mw}\,\rangle$ (100 M$_\odot$/Myr)} \\
  \cline{3-6}
      & log(age/yr)   & MIST                       &   PARSEC                   &   SPOTS                    &   ALL                     \\
  \hline
      & 5.0 -- 7.0  &  $3.99_{-0.32}^{+0.37}$   &   $3.29_{-0.37}^{+0.41}$   &   $3.02_{-0.31}^{+0.38}$   &   $3.42_{-0.33}^{+0.38}$      \\
 20\% & 5.6 -- 6.8  &  $4.23_{-0.38}^{+0.43}$   &   $3.60_{-0.43}^{+0.47}$   &   $3.40_{-0.34}^{+0.40}$   &   $3.75_{-0.38}^{+0.43}$      \\
      & 6.0 -- 6.6  &  $4.78_{-0.44}^{+0.47}$   &   $3.96_{-0.52}^{+0.53}$   &   $2.97_{-0.41}^{+0.45}$   &   $4.05_{-0.45}^{+0.47}$      \\
  \hline
      & 5.0 -- 7.0  &  $5.02_{-0.31}^{+0.36}$   &   $3.87_{0.44}^{0.48}$   &   $3.77_{0.31}^{0.37}$   &   $4.22_{0.35}^{0.40}$     \\
 40\% & 5.6 -- 6.8  &  $5.35_{-0.38}^{+0.44}$   &   $4.29_{0.51}^{0.55}$   &   $4.32_{0.36}^{0.42}$   &   $4.65_{0.24}^{0.29}$     \\
      & 6.0 -- 6.6  &  $4.55_{-0.66}^{+0.71}$   &   $5.06_{0.74}^{0.76}$   &   $3.68_{0.56}^{0.62}$   &   $4.43_{0.41}^{0.44}$     \\
  \hline
\end{tabular}
\end{center}
\end{table}

Fig.~\ref{fig:sfh_results} shows the inferred Orion A SFH using the MIST, PARSEC and SPOTS isochrones. For this derivation of the SFH we limit the \citetalias{mck19} sample to stars with apparent $G$ magnitude brighter than $G_{\rm lim}$\,=\,16.5 mag \textbf{(1275 stars)}. This value was tested to be a realistic completeness limit comparing the data with our simulations in Appendix\,\ref{app:basis}. The median values and error bars (credibility intervals defined in Section\,\ref{sec:nummass}) are listed in columns \ref{fig:sfh_results}\,{\it (a,b,c)} \textbf{and \ref{fig:sfh_results}\,{\it (d,e,f)}} of Table\,\ref{tab:resFig6}. The SFHs from Fig.\,\ref{fig:sfh_results} are plotted next to each other in Fig.\,\ref{fig:sfh_summary}, slightly shifted in log age for easy comparison.

We do not consider age bins $\log t < 5.2$.
There are two reasons for this. First, we do not
have ``young'' enough isochrones to properly assign ages to stars. Second, the meaning of
apparently very young ages is unclear. As discussed further in \S \ref{sec:birthline?},
for timescales comparable to or less than typical estimates of protostellar lifetimes $\sim 0.5$~Myr, the position of the star in the CMD or HD diagram probably reflects initial conditions rather than a true age.
In other words, the properties of the youngest
stars are most likely reflecting variations in
initial conditions of formation rather than the star formation history.

In Table\,\ref{tab:aveFig6} we list the mass-weighted average SFR $\langle \Psi_{mw} \rangle$ for 3 different intervals of log age for the MIST, PARSEC and SPOTS solutions, and all-combined (ALL). The
$\langle \Psi_{mw} \rangle$ values for the ALL case are plotted as {\it gray} colour bands in Fig.\,\ref{fig:sfh_summary}. 
Our results from the optical samples show that, \textbf{assuming 20\% of UBS in our statistical model}, the SFH is consistent with a roughly {\it constant} rate for the last 5-10 Myr \textbf{(see {\it panels (a,b,c)} from Fig.\,\ref{fig:sfh_results} and {\it top panel} from Fig.\,\ref{fig:sfh_summary})}. Counting only the optically bright stars, we find
\begin{equation}\label{eq:psiavg}
   \langle \Psi_{mw} \rangle \approx 380 \pm 70\ \msun\,{\rm Myr}^{-1},
\end{equation}
\noindent
but this value is an underestimate because many stars that are heavily reddened or embedded in nebulosity are not represented in the samples, and could be low by a factor of two or so (see discussion in Sec.~\ref{sec:sfh}).

In appendix \ref{app:ub}, we provide details on the multiplicity frequency and orbital period distribution for our mock binary pairs based on the studies by \citet{duchene13} and \citet{moe2017}. The latter reported a maximum at 50 AU in their orbital separation distribution for solar-type stars. Considering this information and assuming a mean angular resolution of 0.1 arcsecs, we found that about 12\% of the total number of stars in the mock catalogs are UBS. However, \citet{duchene18} found a higher fraction of binary pairs in the ONC with orbital separations from 10 to 60 AU (0.025 to 0.15 arcsecs). As the Gaia EDR3 completeness drops for angular separations below 0.15 arcsecs \citep{fabri21}, a higher fraction of UBS in the Orion A region is plausible. In {\it panels (d,e,f)} of Fig.\,\ref{fig:sfh_results}, we show the SFH inferred for the MK19 sample assuming a 40\% fraction of UBS. The higher UBS fraction translates into a {\it younger} SFH, as expected, but the SFR distribution is similar to that for the case with 20\% UBS.
Fig.\,\ref{fig:sfh_summary} compares the results for the different sets of isochrones. In Table\,\ref{tab:aveFig6} we list the mean SFR for three different intervals of log age.

\section{Discussion}

\subsection{Star formation history}\label{sec:sfh}

The optical samples are significantly
incomplete due to regions of high extinction and nebulosity.
What we may hope is that our samples are representative of
the {\bf relative} SFR as a function of time. Our results for
the different samples are similar, which seems reasonable but of course not conclusive. We note that of the full \citetalias{mck19} sample of all of the know members, consisting of of 5988 stars, only $\sim$50\% have have been selected to good astrometric and photometric data in Gaiato be used in the analysis presented here. The sources that have been excluded are either too extincted (preferentially excluding all of the protostars and many of the disk-bearing stars), or the sources found near the Trapezium due to nebulosity, \citep[which tends to be dominated by the stars with the youngest ages in the region, e.g.,][]{kounkel2021}. 

The CMDs of the various samples
(Fig.~\ref{fig:CMD_samples_2}) show a clear increase in source density at fainter
magnitudes corresponding to older
isochrones. In the absence of significant observational uncertainties,
a relatively steady star formation rate will result in such a pileup in the CMD because pre-main sequence contraction
slows as a star ages \citep{hartmann01}.
A toy model for a completely convective star contracting at roughly constant
temperature (a Hayashi track)
predicts that the stellar luminosity as
\begin{equation}
t - t_0 = (t_{KH}/3) \left [ (L/L_0)^{-3/2} - 1 \right]\,,
\end{equation}
where $L_0$ is the luminosity at the starting
time $t_0$ and $t_{KH} = (3/7) GM^2/R_0$ is the
initial Kelvin-Helmholtz timescale
\citep{hartmann97,hartmann01}.
Thus,
for $t \gg t_0$, the luminosity
decays with age as $L \propto t^{-2/3}$
in reasonable agreement with detailed
calculations. Then changes
$ \Delta \log L$ and magnitude $\Delta m$ are proportional to a change in $\log t$ as the star contracts, 
\begin{equation}
\Delta \log L = -2/3 \,\,  \Delta \log t\,, \hspace{1cm}
    \Delta m = - 5/3 \,\, \Delta \log t\,.
    \label{deltalmag}
\end{equation}

This slowing of contraction is why
the pileup of stars in the CMDs
does not result in a strong spike
in the star formation rate using
either the MIST or the SPOTS
tracks, which are essentially consistent with a constant star formation rate
between $5.5 \lesssim \log t \lesssim 6.7$.
We note in passing that the pileup
of stars in the CMD is generally not
observed in HR diagrams
\citep[see, for instance][]{dario10}.
This is probably due to uncertainties
in extinction corrections, which do not affect the effective temperature if the
latter is derived spectroscopically,
but do affect the stellar colours in such a way as to move (low mass)
stars roughly along isochrones.

As noted above, the optical samples will be especially incomplete for the youngest stars, mainly due to extinction. Our star formation rates for the
MIST and PARSEC isochrones show a substantial drop off at ages 
$\log t < 5.5$ (Figs.~\ref{fig:sfh_results} and \ref{fig:sfh_summary}),
consistent with typical estimates of ages of the heavily-embedded protostar (Class 0 and I) phases \citep{evans09}.

To make a crude estimate of the incompleteness
at the youngest ages, we consider the
{\em Spitzer} and {\em Herschel Space Telescope} surveys of young
stars in Orion \citep{megeath12,fischer17} which identified 
roughly 200 protostars in the area covered by \citetalias{mck19}.
The {\em Spitzer} survey is especially incomplete
in the central region of the ONC. To make a rough
estimate of this, we add the 75 X-ray sources
found in the COUP survey that have no (near)-infrared
counterparts \citep{getman05}.
If we adopt a protostellar lifetime 
of $\sim 0.5$~Myr
\citep[e.g,][]{evans09,dunham14} and a mean
mass of $0.5 \msun$,
this leads to a protostellar star formation
rate of $\sim 275 \msun {\rm Myr}^{-1}$,
comparable to the result in Eq.~(\ref{eq:psiavg}).

At the other end of the age range,
star formation in Orion A must have a finite duration,
and other groups in the Orion complex with ages
$\gtrsim 10$~Myr are not associated molecular gas
\citep[e.g.,][]{kounkel18}. 
 As they are generally not affected strongly by extinction, there should not be significant systematic incompleteness among the older sources outside of the lowest mass stars with intrinsic colours close to the magnitude limit. However, there may be contamination from the other marginally older parts of the Orion Complex along the similar line of sight. As such, the decrease at $\log t \gtrsim 6.7$ is most likely real, reflective of the onset of earliest epoch of star formation in Orion A.

 It is worth noting that while we find an extended period
 of star formation in Orion A, the spatial distribution
 of the stars changes with age, with the youngest stars
 most centrally-concentrated \citep[e.g.,][]{beccari17}.

\subsection{Three populations in Orion?}

From an OmegaCAM survey of a large region in the direction of Orion,
B17 found three bands of 
stars in their CMD that they attributed
to three episodes of star formation.
Although the displacement in $r-i$
from a ridge line of the densest distribution of stars was comparable
to that expected for unresolved binaries and triples,
these authors argued that an explanation solely in terms of multiple systems could not reproduce the observations, 
as they would require an unlikely companion mass ratio skewed toward
equal masses. 
In a followup paper, J19
used proper motion and parallaxes from Gaia to select a subset of their OmegaCAM sample, finding the three populations as before. 

The analysis of our samples do not show
evidence for three populations of stars with distinctly
different ages, but rather a much more constant,
or at least continuous, rate of star formation.
{\color{blue} (See the discussion in Appendix A8 for testing the
suggested B17 and J19 bursts of star formation.)}
The origin of the discrepancy is not clear.
It seems unlikely that the difference is due to photometry,
although B17 and J19 use OmegaCam rather than Gaia magnitudes;
although it is worth noting that the magnitude range spanned
by the samples differ.  This does not seem to
be a question of statistics,
as the numbers of stars in the
samples are roughly the same, as
the J19 sample size of 852 stars is similar to the K18 and half
of the \citetalias{mck19} samples analyzed here.
\textbf{
A more probably explanation for the discrepancy can be seen in Figure 3, where our Gaia samples miss a significant amount of stars in the ONC region, possibly hiding the characteristic footprints produced by an episodic SFH. In contrast, the OmegaCAM photometric sample used in B17 and J19 has a higher completeness level and more accurate photometry than Gaia around the ONC center.
}
Further investigation of the differences is warranted.

\subsection{Is there a birthline?}

\label{sec:birthline?}

From at least the time of \cite{larson69} it has been clear that stars do not form with arbitrarily large radii,
basically because contraction timescales cannot be much shorter than the time it takes for a star to accrete its mass. 
\cite{stahler83,stahler88} 
made the first systematic efforts to refine
the starting radii of young stars, showing that under certain assumptions,
low-mass protostars would evolve along a mass-radius relation
that was regulated by deuterium fusion. This theory
predicts a fairly well-defined locus of initial
radii at the end of protostellar accretion, corresponding
to a ``birthline'' in the HR diagram where young stars
would first become optically visible. However,
for stars with masses $> 1 \msun$, the energy release
by deuterium becomes less important, and results are
sensitive to the assumed mass infall rate $\mdot_{in}$ \citep{pallastahler90}. 

Subsequent theoretical work has clouded this picture.
Birthline positions
are sensitive to the assumed initial mass and radius of the core formed in the early hydrodynamic collapse
\citep{hartmann97,hosokawa11,baraffe12}, and this in principle
could be sensitive to $\mdot_{in}$.
Moreover, major uncertainties arise from the likely need to
go beyond formation by radial infall and include accretion from the circumstellar disks that are an inevitable result of rotating protostellar
cloud collapse. The amount of thermal
energy that gets added to the protostar via
disk accretion is uncertain;
so far, evolutionary calculations have simply parameterized
heat addition, with considerably varying results for protostellar 
radii \citep{hartmann97,baraffe09,baraffe12,hosokawa11}.

Standard evolutionary tracks, such as those of MIST and PARSEC,
indicate that in the mass range probed by our samples the
onset of D fusion occurs at an age $t \sim 0.3-0.5$~Myr.
We do not detect a pileup of stars along the 
D main sequence using the MIST and PARSEC isochrones,
although as discussed above we are substantially
incomplete at younger ages.

To explore this further, in panel {\it (c)} of Fig.~\ref{fig:CMDt} we show the SPOTS isochrones with the \citetalias{mck19} sample, including high proper motion stars (see their figure 1), many of which populate the CMD high above the youngest isochrone (this is also true for the MIST and PARSEC isochrones in Fig.~\ref{fig:CMDt}). These stars may well have had a special origin, involving ejection from a multiple system or interactions with other stars in the densest region of the ONC, but their existence seems to be strong evidence that not all objects begin their pre-main sequence contraction from a common birthline.

The age where D fusion becomes
important in cool stars is systematically older in the SPOTS tracks compared to the non-magnetic MIST and PARSEC tracks. This seems to be a joint effect of isochrone initial conditions and
differences in internal structure caused by star spots.
None of the isochrone grids properly account for the effects of pre-main sequence accretion, and variations in the initialization conditions of the models are a genuine uncertainty in the models which contribute to this observation. However, we find that star spots generically lengthen the Kelvin-Helmholtz timescale as the result of flux redistribution in surface convection zones and the subsequent change in stellar interiors \citep{somers15}. Between SPOTS non-spotted and spotted models, this effect lengthens the D fusion timescale due to the structural effects of stellar magnetism. The longer D fusion timescale means that more stars lie above the D fusion main sequence in magnetic tracks, making the effect of deuterium burning even less important in establishing a birthline.

\section{Conclusions}

We have developed statistical methods within a Bayesian framework to infer the SFH from Gaia photometric surveys of pre-main sequence populations. We applied corrections for extinction on SFH inference using a subsample with measured $A_V$ to develop a statistical relation between age and reddening. We also implemented a statistical correction of the bias in the SFH introduced by unresolved binary stars, which we find has only a small effect.

Using the MIST and PARSEC evolutionary tracks based on standard assumptions, we find that star formation in Orion A has proceeded at a relatively constant rate between ages of about 0.3 and 5 Myr, in contrast to other studies suggesting multiple epochs of star formation. The SPOTS tracks suggest a similarly constant rate extending to 10 Myr. We find no evidence for a well-constrained deuterium-burning ``birthline'' in the Gaia colour-magnitude diagram; this is an especially strong result if the SPOTS tracks apply.

The method's hierarchical structure allowed us to transition from individual photometry and parallaxes to a population attribute like its SFH.
Within this general arrangement, we have the option to analyze different photometric data or include prior parameters, e.g. astrometric quantities, opening the possibility to infer star formation histories from photometric surveys of other regions.

\section*{Acknowledgements}

JAA and GB acknowledge financial support from the National Autonomous University of M\'exico (UNAM) through grants DGAPA/PAPIIT IG100319 and BG100622.
JAA acknowledges support from the IRyA (UNAM) computer department and for granting him the required CPU time in the computers acquired through CONACyT grant CB2015-252364.
JAA also acknowledge to CONACyT for the financial support CB-A1-S-25070.
MK acknowledges support from the Vanderbilt initiative in data intensive astrophysics (VIDA).
LC acknowledges support from TESS Cycle 5 GI program G05113 and NASA grant 80NSSC19K0597.

This work has made use of data from the European Space Agency (ESA) mission
{\it Gaia} (\url{https://www.cosmos.esa.int/gaia}), processed by the {\it Gaia}
Data Processing and Analysis Consortium (DPAC,
\url{https://www.cosmos.esa.int/web/gaia/dpac/consortium}). Funding for the DPAC
has been provided by national institutions, in particular the institutions
participating in the {\it Gaia} Multilateral Agreement.

\section*{Data availability}

This research or product makes use of public auxiliary data provided by ESA/Gaia/DPAC/CU5 and prepared by Carine Babusiaux. The SPOTS models are publicly available and can be downloaded at \url{https://zenodo.org/record/3593339}.

\bibliographystyle{mnras}
\bibliography{references}

\appendix

\section{Mock catalogue}\label{app:basis}

In this section we build a mock Orion A stellar population to test the degree to which our statistical model (Sec.~\ref{sec:inference}) can recover its SFH. 
To build the mock catalogue we need to specify:
{\it (a)} a set of isochrones covering the relevant ages and metallicity;
{\it (b)} the photometric properties of the stars along the isochrones in the relevant bands;
{\it (c)} the time dependence of the SFR and the amount of mass formed into stars in each age bin;
{\it (d)} the IMF;
{\it (e)} the fraction of stars in binary systems as a function of mass or spectral type; 
{\it (f)} the extinction by dust along the line of sight; and
{\it (g)} the radial distance distribution of the stars in the cluster.
In the case of Orion A, we assume that all the stars are at the same distance since the depth of the cluster is much less than its distance from us (400 pc, $dm$~=~8 mag; for details on 3D simulations see \citetalias{alzate21}).
These properties must be expressed as functions that can be sampled stochastically.
We follow the procedure described in \citetalias{alzate21} with some modifications, summarized below for the benefit of the reader. 

\subsection{Isochrones}\label{app:iso}

We use the MIST, PARSEC, and SPOTS isochrones for solar metallicity listed in Table~\ref{tab:iso}. For the simulations we use a larger number of isochrones than for
the statistical model. 
From the MIST database we select 41 isochrones ranging from log age(yr) 5 to 7 in steps of 0.05. Isochrones below 0.1 Myr are not available in the MIST database.
From the PARSEC database we use 52 isochrones ranging from log age(yr) 4 to 7 in steps of 0.1 from 4 to 4.9 and 0.05 from 4.9 to 7.00. 
From the SPOTS database we use 61 isochrones ranging from log age(yr) 4 to 7 in steps of 0.05.
In Table~\ref{tab:imass} we indicate the upper mass limit  $M_{up}$ for each isochrone. For all the isochrones in use, the lower mass limit is $M_{low}$\,=\,0.1\,$\msun$. 

\begin{table*}
  \begin{center}
  \caption{\label{tab:imass}Upper mass limit $M_{up}$ for different isochrone sets and incompleteness correction from simulations for constant SFR using the \citet{Kroupa2001} IMF.}
  \begin{tabular}{ccrcccccccccccc}
         && \mlc{3}{c}{Isochrone mass limit$^a$} &&& \mlc{8}{c}{Incompleteness correction $\epsilon_{8.5}$}  \\
  \cline{3-5}\cline{8-15}
         && \mlc{3}{c}{$M_{up}\ (\msun)$} &&& \mlc{3}{c}{Single Stars}   &&& \mlc{3}{c}{Single Stars + UBS} \\
  \cline{3-5}\cline{8-10}\cline{13-15}
  $\log({\rm age}/{\rm yr})$ && MIST & PARSEC & SPOTS &&& MIST  & PARSEC & SPOTS  &&& MIST  & PARSEC & SPOTS  \\
  \cline{1-1}\cline{3-5}\cline{8-10}\cline{13-15}
   4.0   &&         &   12  &   1.3     &&&          &  1.06  &  1.56  &&&       &  1.15  &  1.72  \\
   4.2   &&         &   12  &   1.3     &&&          &  1.17  &  1.44  &&&       &  1.29  &  1.33  \\
   4.4   &&         &   12  &   1.3     &&&          &  1.15  &  1.54  &&&       &  1.15  &  1.49  \\
   4.6   &&         &   12  &   1.3     &&&          &  1.23  &  1.47  &&&       &  1.30  &  1.52  \\
   4.8   &&         &   12  &   1.3     &&&          &  1.28  &  1.65  &&&       &  1.29  &  1.62  \\
   5.0   &&   296   &   12  &   1.3     &&&    1.62  &  1.34  &  1.60  &&& 1.63  &  1.37  &  1.81  \\
   5.2   &&   262   &   11  &   1.3     &&&    1.66  &  1.49  &  1.83  &&& 1.68  &  1.59  &  1.70  \\
   5.4   &&   274   &   11  &   1.3     &&&    1.76  &  1.62  &  1.88  &&& 1.71  &  1.60  &  1.93  \\
   5.6   &&   280   &   11  &   1.3     &&&    1.83  &  1.64  &  2.14  &&& 1.87  &  1.70  &  2.10  \\
   5.8   &&   285   &   11  &   1.3     &&&    1.97  &  1.78  &  2.47  &&& 2.03  &  1.80  &  2.36  \\
   6.0   &&   288   &   11  &   1.3     &&&    2.17  &  1.88  &  2.66  &&& 2.14  &  1.82  &  2.51  \\
   6.2   &&   295   &   11  &   1.3     &&&    2.27  &  1.89  &  2.50  &&& 2.26  &  1.89  &  2.47  \\
   6.4   &&   292   &   11  &   1.3     &&&    2.52  &  2.18  &  2.73  &&& 2.59  &  2.18  &  2.63  \\
   6.6   &&    73   &   12  &   1.3     &&&    2.92  &  2.47  &  3.24  &&& 2.90  &  2.44  &  3.09  \\
   6.8   &&    33   &   11  &   1.3     &&&    3.27  &  2.73  &  4.08  &&& 3.27  &  2.71  &  3.84  \\
   7.0   &&    20   &   12  &   1.3     &&&    3.57  &  2.95  &  5.08  &&& 3.54  &  2.92  &  4.66  \\
  \cline{1-1}\cline{3-5}\cline{8-10}\cline{13-15}
  && \multicolumn{3}{l}{$^aM_{low}$\,=\,0.1\,$\msun$\ for all sets.}
  \end{tabular}
  \end{center}
\end{table*}

\subsection{Stochastic sampling of the SFR}\label{app:sfr}

For a constant SFR from $t$~=~$t_1$ to $t$~=~$t_2$, we use Eq. (B6) from \citetalias{alzate21},
\begin{equation}
t(T)= t_2 - T(t_2 - t_1)\ \ \ \ \ \ \ \ \ \ \ \ \ \ \ 0 \le T \le 1,
\label{csfr2}
\end{equation}
to select the age $t$ of any given star created stochastically according to 
the randomly sampled variable $T$.

\subsection{Stochastic sampling of the IMF}\label{app:imf}

\begin{flushleft}
{\it (a) Single power law IMF}
\end{flushleft}
\noindent
For a single power law IMF, e.g., \citet{salp55},
\begin{equation}
\Phi(m) = dN/dm = Cm^{-(1+x)},
\label{phi}
\end{equation}
we can write
\begin{equation}
m(N) = [(1-N)m_l^{-x} + Nm_u^{-x}]^{-\frac{1}{x}}\ \ \ \ \ \ \ \ \ \ \ \ \ \ \ 0 \le N \le 1,
\label{nc}
\end{equation}
cf. \citet{sf97} and Eq.~(B13) from \citetalias{alzate21}.

\medskip
\begin{flushleft}
{\it (b) Double power law IMF}
\end{flushleft}
\noindent

In the case of a two-segment power law IMF, e.g., \citet{Kroupa2001}, written in general as,
\begin{equation}
\Phi(m) = 
\begin{cases}
    C_1\ m^{-(1+x_1)}\ \ \ {\rm if}\ \ \ m_l \le m \le m_c\cr
    C_2\ m^{-(1+x_2)}\ \ \ {\rm if}\ \ \ m_c \le m \le m_u,
\end{cases}
\label{phi2}
\end{equation}
we have (Eq. B20 from \citetalias{alzate21}),
\begin{equation}
m(N) = \begin{cases} 
       [ \frac{(N_c-N)m_l^{-x_1} + Nm_c^{-x_1}}{N_c} ]  ^{-\frac{1}{x_1} }\ \ \ \ \ \ \ \ \ \ \ \ \ \ \ {\rm if}\ \ \ N \le N_c \cr
       [ \frac{(1 - N)m_c^{-x_2} + (N-N_c)m_u^{-x_2} }{1 - N_c } ]  ^{-\frac{1}{x_2} } \ \ \ \ \ \ \ \ \ {\rm if}\ \ \ N > N_c,
       \end{cases}
\label{norm2}
\end{equation}
where again $0 \le N \le 1$.

\medskip
Sampling $N$ with a random number generator we obtain from Eqs.~(\ref{nc}) and (\ref{norm2}) values of $m$ that follow the IMFs in Eqs.~(\ref{phi}) and (\ref{phi2}), respectively.
In these equations, $(m_l,m_u)$~=~$(0.1,100)$ M$_\odot$ are the lower and upper mass limits of star formation.
For the \citet{salp55} IMF, $x$~=~1.35.
For the \citet{Kroupa2001} universal IMF, $m_c$~=~0.5 M$_\odot$, $(x_1,x_2)$~=~(0.3,1.3), and $N_c = 0.72916$, indicating that 72.9\% of the stars are born with $m \le 0.5~M_\odot$.

\subsection{Binary stars}\label{app:ub}

\begin{table}
\caption{\label{tab:dk}Multiplicity frequency from \citet[][Table 1]{duchene13}}
\centering
\begin{tabular}{cc}
\hline
 Mass Range & MF \\
\hline
M$_* < 0.1$\ M$_\odot$            & 22\% \\
$0.1 \leq$ M$_* < 0.6$\ M$_\odot$ & 26\% \\
$0.6 \leq$ M$_* < 1.4$\ M$_\odot$ & 44\% \\
$1.4 \leq$ M$_* < 8$\   M$_\odot$ & 50\% \\
$8   \leq$ M$_* < 16$\  M$_\odot$ & 60\% \\
$M_* \geq 16$\          M$_\odot$ & 80\% \\
\hline
\end{tabular}
\end{table}

From \citet[][]{duchene13} we adopt the multiplicity frequency (MF) for stars of different mass listed in Table~\ref{tab:dk}. This is a simplified version of their table 1 that we judge sufficient for the goals of this simulation. To decide if a star of mass $m$ is part of a binary system, we draw a random number $B$. If
\begin{equation}
B \le \frac{{\rm MF}(m)}{100} \ \ \ \ \ \ \ \ \ \ \ \ \ \ \ \ \ \ \ \ \ \ 0 \le B \le 1,
\label{eq:dk}
\end{equation}
the star is assumed to be the primary of a binary system. The mass $m_2$ of the secondary follows from \citet[][p. 20 and fig. 2]{duchene13}, namely
\begin{equation}
q = \frac{m_2}{m_1} = 
\begin{cases}
    0.1 + 0.9*Q\ \ \ \ \ \ \ {\rm if}\ \ \ m_1 \ge 0.3 {\rm M}_\odot\cr
    0.5 + 0.5*Q\ \ \ \ \ \ \ {\rm if}\ \ \ m_1 < 0.3 {\rm M}_\odot,
\end{cases}
\label{qrat}
\end{equation}
where $0 \le Q \le 1$ is a random number, and $q \le 1$. Eq.~(\ref{qrat}) assumes that $q$ is uniformly distributed 
between 0.1 and 1 if $m_1 \ge 0.3$ M$_\odot$ and 
between 0.5 and 1 if $m_1 < 0.3$ M$_\odot$. 

We assign an orbital period to each binary pair following \citet{moe2017}, who find that the orbital periods of solar type MS binaries fulfill
\begin{equation}\label{eq:logp}
{\rm log}_{10}\ P({\rm days}) \sim \mathcal{N}(4.9,2.3,Y),
\end{equation}
where $\mathcal{N}(4.9,2.3)$ is a normal distribution with
$\langle{\rm log}~P({\rm days})\rangle$ = 4.9 and $\sigma_{{\rm log}~P({\rm days})}$~=~2.3,
and $Y$ is a random number needed to sample $\mathcal{N}$ stochastically.

\medskip
The semi-major axis of the orbit then follows from Kepler's law
\begin{equation}\label{eq:aau}
{\rm log}~a({\rm au})\,=\,\frac{1}{3}\,{\rm log}\,(m_1+m_2)\,+\,\frac{2}{3}\,{\rm log}\,P({\rm yr}).
\end{equation}

\medskip
Since the \g resolution limit is $0.1''$, then if
\begin{equation}
\theta = \frac{a({\rm au})}{d({\rm pc})}
\begin{cases}
< 0.1''\ \ \ \ \ \ \ \Rightarrow {\it unresolved}\ {\rm binary\ pair}\cr
\ge 0.1''\ \ \ \ \ \ \ \Rightarrow {\it resolved}\ {\rm binary\ pair},
\end{cases}
\label{arat}
\end{equation}
where $d$~=~400\, pc is the approximate distance to Orion A. 

\subsection{Stellar photometry}\label{app:phot}

Once $t(T)$ and $m(N)$ are available from Sections~\ref{app:sfr}, \ref{app:imf} and \ref{app:ub}, we search for the isochrones of age $t^{-}$ and $t^{+}$ bracketing age $t$ and, by interpolation, find the flux in the \g $G,\ G_{\rm BP}$ and $G_{\rm RP}$ bands corresponding to a star of mass $m$ in each of these two isochrones. These fluxes are them interpolated logarithmically in $t$ to the required age. Resolved binaries are treated in the simulation as two coeval single stars. The fluxes of unresolved binaries are added before including the pair as a single star in the simulation.

\subsection{Extinction by dust}\label{app:dust}

To assign a value of $\av$ to each star in our simulations we use the normalized cumulative distributions $D(\av)$ shown in the right hand side panels of Fig.~\ref{fig:av}. Sampling $D(\av)$ with a random number generator, we can read from the abscissa the corresponding value of $\av$ according to the star age.
Then we follow the procedure explained in Sec.~\ref{sec:edust} to compute the extinction in the \g  $G,\ G_{\rm BP}$ and $G_{\rm RP}$ bands.

\subsection{Catalogues}\label{app:cat}

For every star in our simulated catalogue, we need to generate a set of 6 random numbers $(T,N,B,Q,Y,D)$ which determine the star age, mass, multiplicity, mass ratio, orbital period and dust extinction, respectively. This is performed in a loop that is stopped once the desired number of stars or cluster mass is achieved.

\subsubsection{Incompleteness Correction}\label{app:incompcorr}
\begin{figure*}
\begin{center}
    \includegraphics[width=\textwidth]{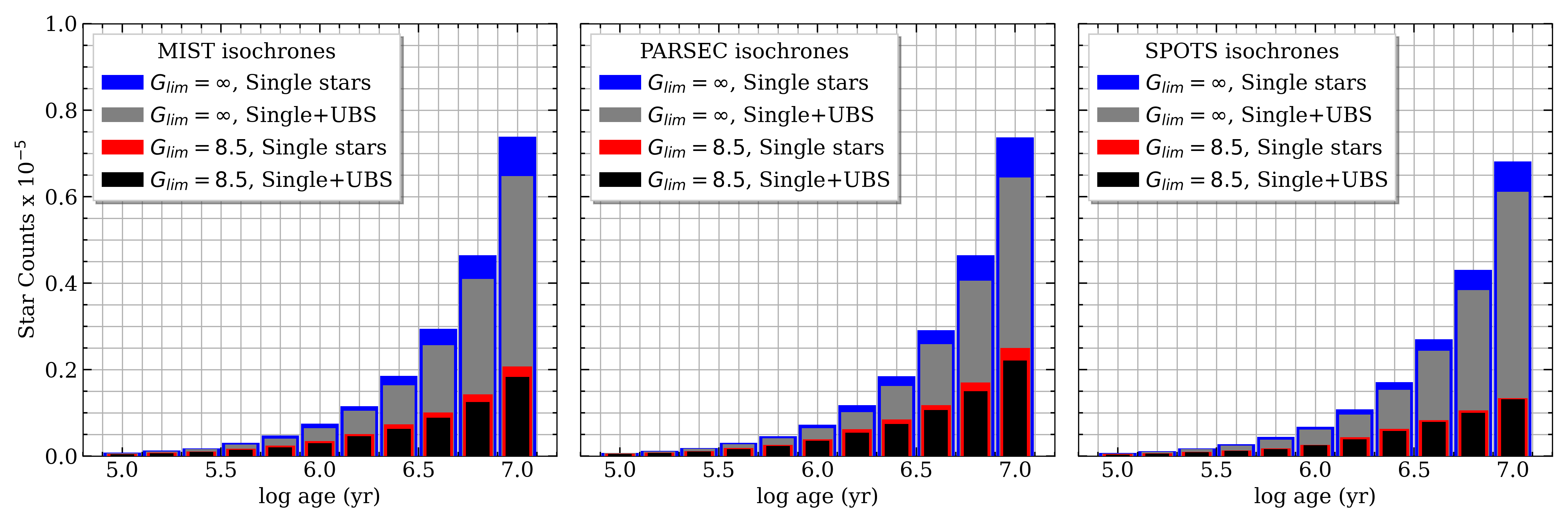}
    \caption{$N_{*}^{i}$ for simulated Orion A populations of 200,000 stars according to the MIST ({\it left}), PARSEC ({\it center}) and SPOTS ({\it right}) isochrone sets assuming a constant SFR.
    The {\it blue} and {\it gray} histograms include all the stars in the simulation. 
    The {\it red} and {\it black} histograms include only the stars brighter than $G$\,=\,8.5 mag.
    The {\it blue} and {\it red} histograms correspond to single star models, whereas the
    the {\it gray} and {\it black} histograms correspond to models which include both single stars and UBS.
    The resulting incompleteness corrections, defined in Eq.~(\ref{eq:incomp}), are listed in Table~\ref{tab:imass} and plotted in Fig.~\ref{fig:incompleteness}.
    \label{fig:ic_histo}}
\end{center}
\end{figure*}

\begin{figure}
\begin{center}
    \includegraphics[width=1.1\columnwidth]{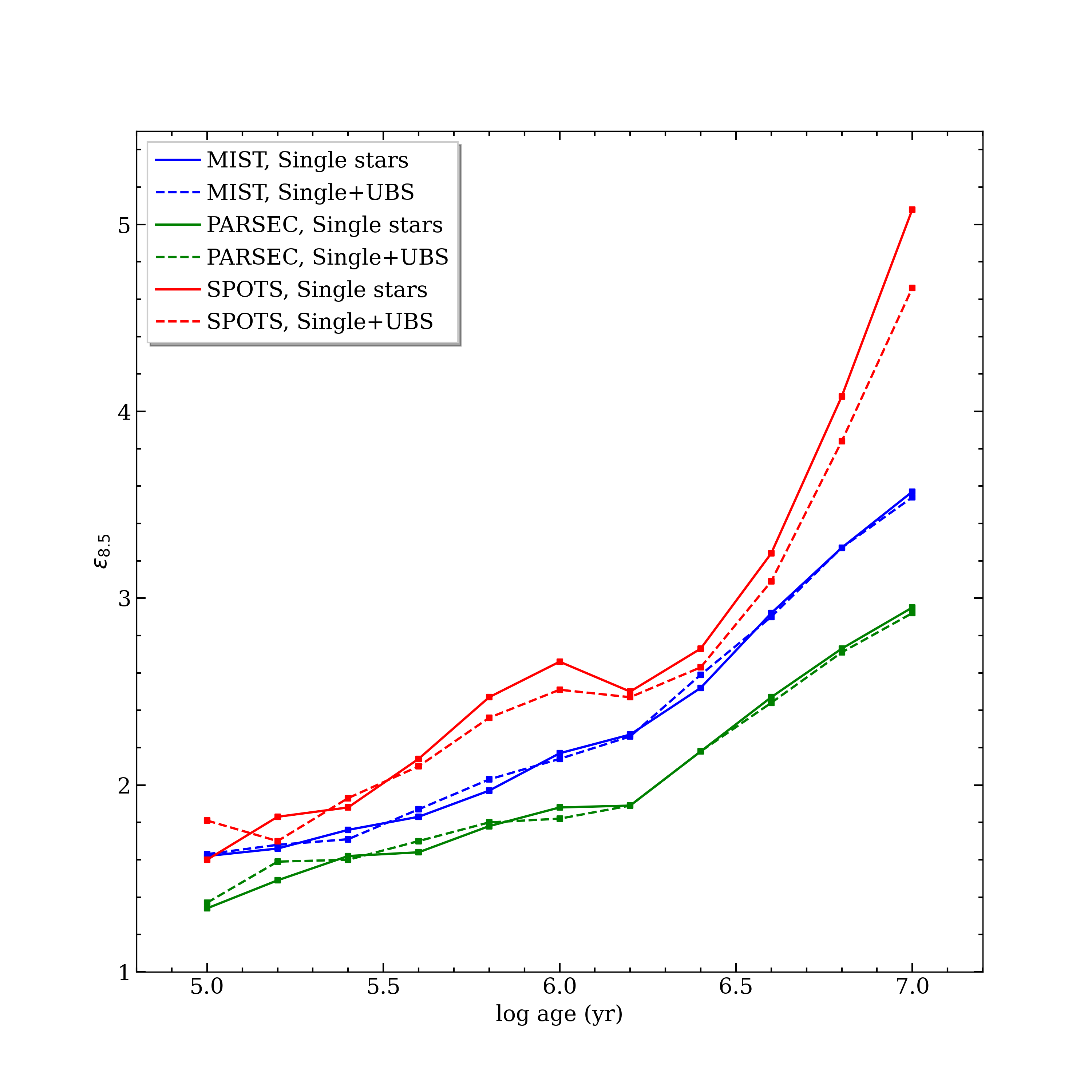}
    \caption{Incompleteness correction from simulation with constant SFR (Fig.~\ref{fig:ic_histo} and Table\,\ref{tab:imass}) for the \citet{Kroupa2001} IMF.
    \label{fig:incompleteness}}
\end{center}
\end{figure}

First, we use our simulations to determine the incompleteness correction to be applied to our statistical inference of the number of stars in a given population. For this purpose we simulate a population of 200,000 single stars with the properties of Orion A. Fig.~\ref{fig:ic_histo} shows histograms of the number of stars per bin of log age for six sets of models. We define the incompleteness correction as
\begin{equation}\label{eq:incomp}
\epsilon_{8.5}(t_i) = \frac{N_{*}^{i}(G_{\rm lim}=\infty)} {N_{*}^{i}(G_{\rm lim}=8.5)},
\end{equation}
where $N_{*}^{i}(G_{\rm lim}$\,=\,$\infty$) and $N_{*}^{i}(G_{\rm lim}$\,=\,8.5) are the number of stars to the faintest $G$ magnitude and to $G$\,=\,8.5 mag, respectively. The resulting $\epsilon_{8.5}(t_i)$ for the six models are listed in Table\,\ref{tab:imass} and plotted in Fig.\,\ref{fig:incompleteness}.
To test the dependence of $\epsilon_{8.5}$ on $\Psi(t)$, we repeated our simulations for a variety of SFR's. We conclude that for fixed IMF and $p(A_V)$, $\epsilon_{8.5}$ is insensitive to the presence of UBS in the simulated population and to the time behaviour of its SFR.

\subsubsection{Orion A simulations}\label{app:oriona}

\begin{figure*}
    \centering
    \includegraphics[width=0.98\textwidth]{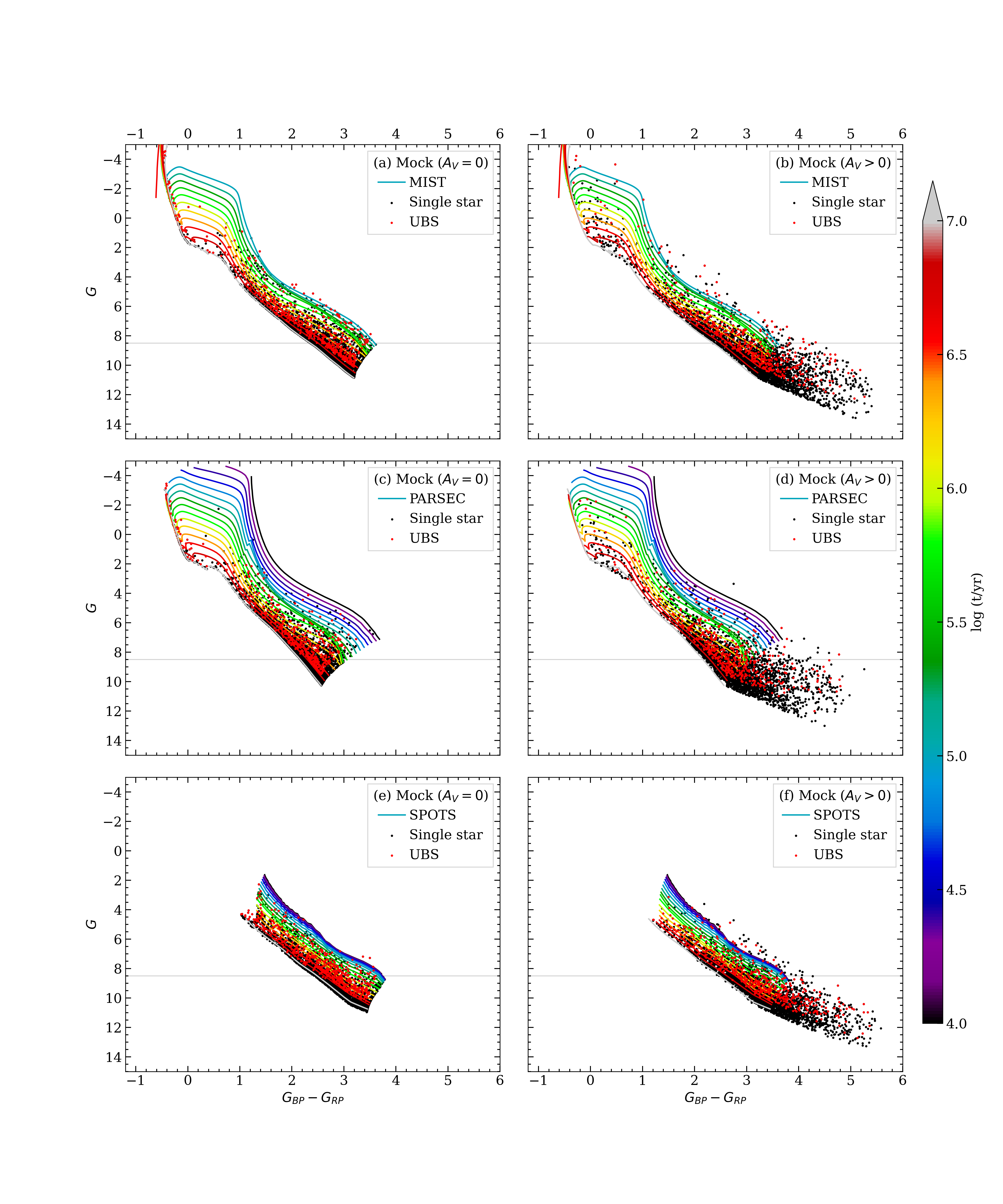}
    \caption{CMD diagrams showing simulated populations of 4000 stars using the MIST, PARSEC and SPOTS isochrones, ignoring extinction by dust ($\av$\,=\,0, {\it panels a,c,e)} and modelling extinction as indicated in Sec.~\ref{app:dust} {\it (panels b,d,f)}. 
    Single and resolved binary stars are indicated by {\it black dots} and UBS by {\it red dots}. In all panels, isochrones ranging in log age (yr) from 4 to 7 in steps of 0.2 are plotted as lines, colour coded by age as indicated in the colour bar. For the MIST data set the 5 youngest isochrones are not available. The isochrones span the mass range indicated in Table~\ref{tab:imass}. A horizontal line is drawn at absolute magnitude $G=8.5$, corresponding to a limiting apparent magnitude $G_{lim}=16.5$ at 400 pc.}
    \label{fig:CMDtt}
\end{figure*}

\begin{figure*}
    \centering
    \includegraphics[width=0.8\textwidth]{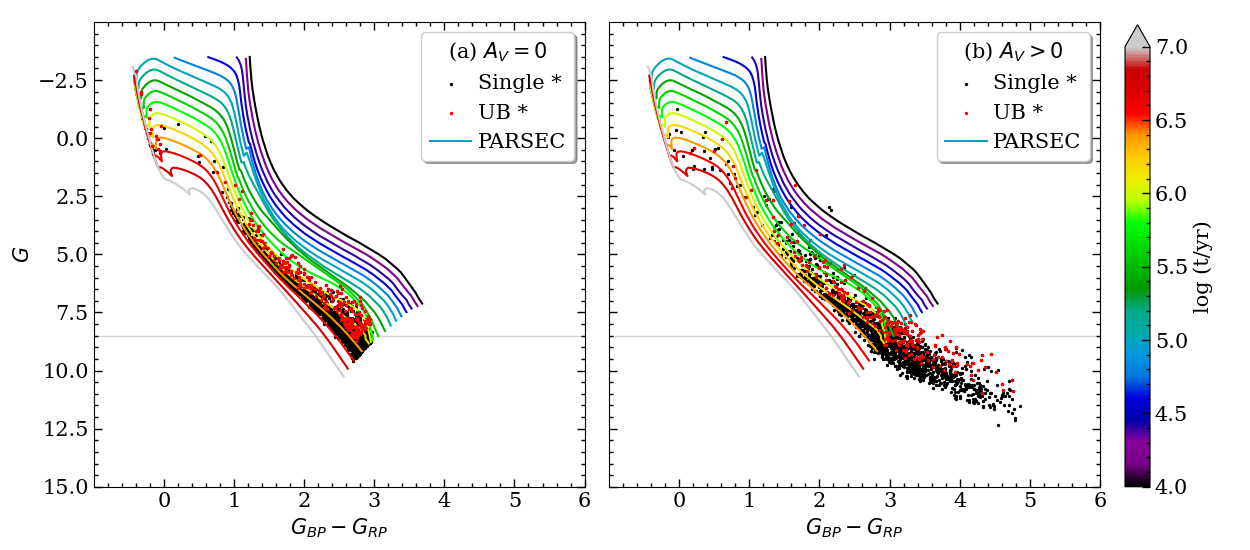}
    \caption{
      CMDs showing simulated populations of 2000 stars using the PARSEC isochrones and
      the \citet{beccari17} bursty SFH,
      ignoring extinction by dust ($\av$\,=\,0, {\it panel a)} and modelling extinction as indicated in Sec.~\ref{app:dust} {\it (panel b)}. 
      Single and resolved binary stars are indicated by {\it black dots} and UBS by {\it red dots}. In both panels, isochrones ranging in log age (yr) from 4 to 7 in steps of 0.2 are plotted as lines, colour coded by age as indicated in the colour bar. The isochrones span the mass range indicated in Table~\ref{tab:imass}. A horizontal line is drawn at absolute magnitude $G=8.5$, corresponding to a limiting apparent magnitude $G_{lim}=16.5$ at 400 pc. The colour bar spans the same range as in Fig.\,\ref{fig:CMDtt}.
      \label{fig:CMD_becc}
      }
\end{figure*}

We simulate the stellar population of Orion A following the procedure described above assuming a constant $\Psi$\,=\,200$\,{\rm M}_\odot\,{\rm Myr}^{-1}$. The results of one of these simulations for 4000 stars\footnote{This number, found by trial and error, results in age bins with numbers of stars close to the observed ones.} using the MIST, PARSEC and SPOTS isochrone sets are shown as CMDs in Fig.~\ref{fig:CMDtt}, highlighting the age distribution of the stars. 
In Fig.~\ref{fig:CMDtt} single and resolved binary stars are indicated by {\it black dots} and UBS by {\it red dots}. 
Isochrones ranging in log age (yr) from 4 to 7 in steps of 0.2 are plotted as lines, colour coded as indicated in the colour bar.
For the MIST data set the 5 youngest isochrones are not available.
{\it Panels (a,c,e)} show the position of the simulated stars when extinction is ignored ($\av$\,=\,0). {\it Panels (b,d,f)} show the position of the same stars when extinction is modelled as indicated in Sec.~\ref{app:dust}. The horizontal line at absolute magnitude $G=8.5$, corresponds to a limiting apparent magnitude $G_{lim}=16.5$ at 400 pc.

Additionally, we build a synthetic catalog for stars born in three different Gaussian bursts, as prescribed by \citet{beccari17} for the ONC, with mean age at 2.87, 1.88, 1.24 Myr in lookback time, and standard deviation 0.385,  0.370, and 0.225 Myr, respectively. We use the PARSEC isochrones to simulate populations of 2000 stars. The resulting CMD for one of these simulations  is shown in Fig.\,\ref{fig:CMD_becc}.

\subsubsection{Binary frequency}\label{app:binaryfreq}

\begin{table*}
\begin{center}
 \caption{\label{tab:tmodel}Different versions of our hierarchical model.
 }
 \begin{tabular}{lccccl}
 \hline
Model  & Single & Unresolved & \multirow{2}{*}{Extinction} & Incompleteness & Model \\
ID     & Stars  & Binaries   &                             & Correction$^a$ & Description\\
 \hline
S       & \checkmark & --         & --         & --         & all stars treated as single, extinction ignored \\
S+UB    & \checkmark & \checkmark & --         & --         & single and binary stars, UBS treated as in Sec.~\ref{sec:ubin}, extinction ignored\\
S+E     & \checkmark & --         & \checkmark & \checkmark & all stars treated as single, extinction modelled according to Sec.~\ref{sec:edust}\\
S+UB+E  & \checkmark & \checkmark & \checkmark & \checkmark & single and binary stars, extinction and UBS treated according to Sec.~\ref{sec:edust} and \ref{sec:ubin}\\
\hline
\multicolumn{5}{l}{$^a$Defined in Sec.~\ref{app:incompcorr}}.
\end{tabular}
\end{center}
\end{table*}

In our simulated catalogue, $\approx$\,47\% of the stars are formed in binary systems, $\approx$\,48\% of these pairs are resolved at the distance of Orion A, resulting in a fraction of $\approx$\,88\% apparently single stars. The remaining $\approx$\,12\% are UBS. 

\subsection{Testing the statistical model}\label{app:tests}

\begin{figure*}
    \centering
    \includegraphics[width=\textwidth]{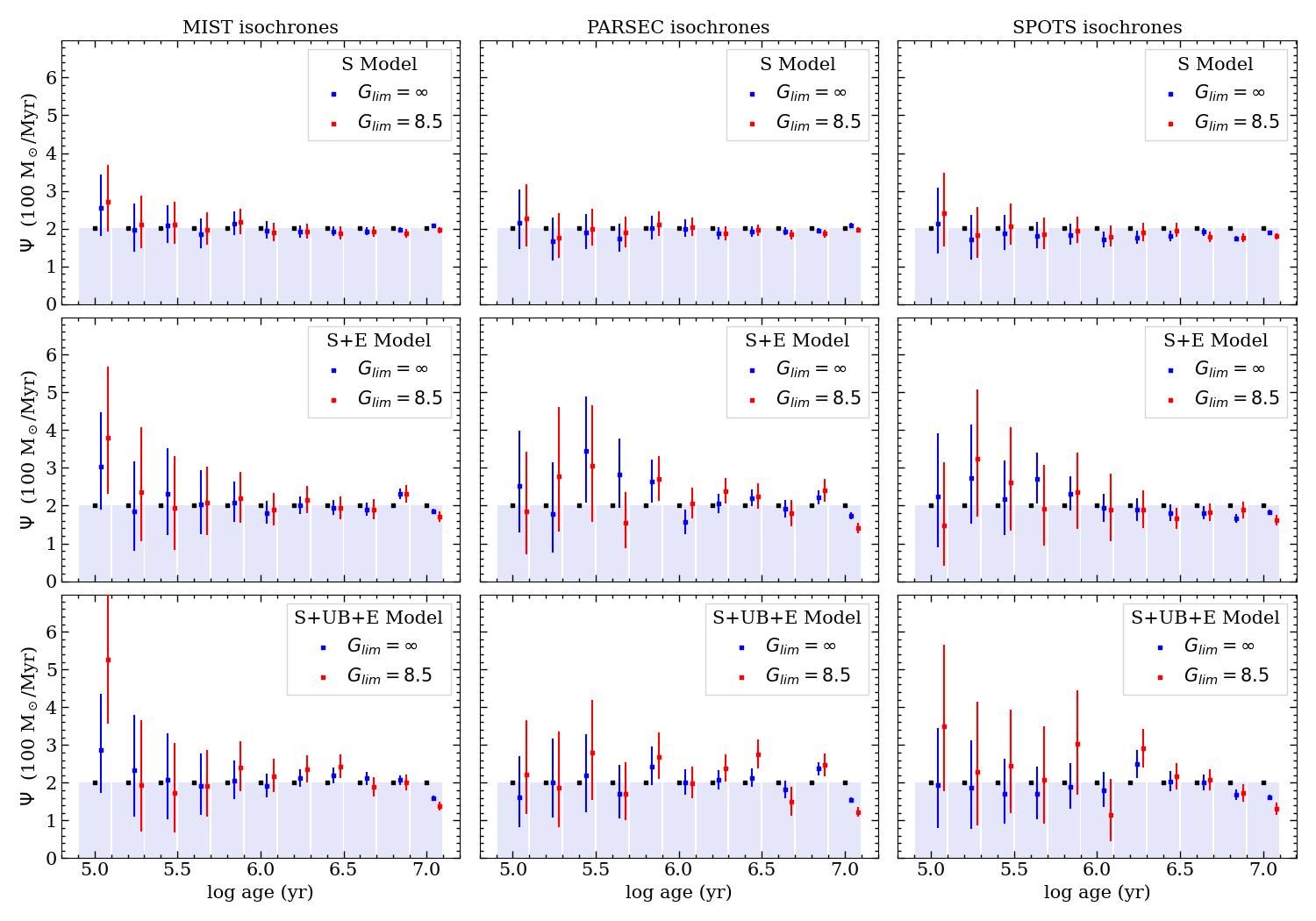}
    \caption{
    Recovered SFH for our simulations using the S {\it (top panels)}, S+E {\it (middle panels)} and S+UB+E models {\it (bottom panels)}, defined in Table~\ref{tab:tmodel}, when we include all the stars in the simulation ($G_{lim}$\,=\,$\infty$) and only the stars brighter than $G_{lim}$\,=\,8.5 mag. The {\it black dots} are drawn at the {\it true} value $\Psi$\,=\,200$\,{\rm M}_\odot\,{\rm Myr}^{-1}$ used in the simulation. We analyzed 10 individual realizations of 4000 stars each. The {\it blue dots} and {\it blue error bars} indicate, respectively, the
    median $\Psi$ and the credibility interval (Sec.\,\ref{sec:nummass}) resulting from the Bayesian inference for the 10 simulations for the $G_{lim}$\,=\,$\infty$ case. The {\it red dots} and {\it red error bars} correspond to the case $G_{lim}$\,=\,8.5. The {\it blue} and {\it red dots} have been shifted slightly in log age for clarity. The {\it light gray} vertical bars are drawn to guide the eye separating the different age bins. Each column corresponds to a different isochrone set, indicated in the {\it top label.}
    \label{fig:hist_all}}
\end{figure*}

\begin{figure*}
    \centering
    \includegraphics[width=\textwidth]{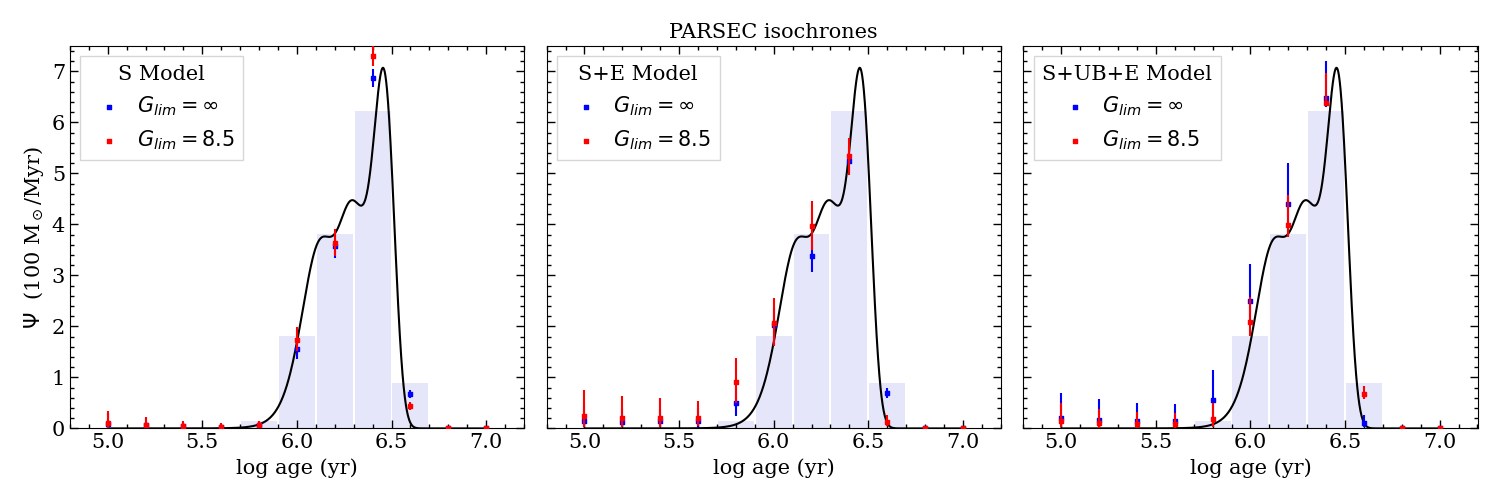}
    \caption{
     Recovered SFH from our simulated catalog using the \citet{beccari17} bursty SFH for the S {\it (left panel)}, S+E {\it (middle panel)} and S+UB+E  {\it (right panel)} models, defined in Table~\ref{tab:tmodel}, when we include all the stars in the simulation ($G_{lim}$\,=\,$\infty$) and only the stars brighter than $G_{lim}$\,=\,8.5 mag. The {\it black line} is drawn following \citet{beccari17} SFH. We analyzed 10 individual realizations of 2000 stars each. The {\it blue dots} and {\it blue error bars} indicate, respectively, the
     median $\Psi$ and the credibility interval (Sec.\,\ref{sec:nummass}) resulting from the Bayesian inference for the 10 simulations for the $G_{lim}$\,=\,$\infty$ case. The {\it red dots} and {\it red error bars} correspond to the case $G_{lim}$\,=\,8.5.
     The {\it light gray} vertical bars are drawn to show how the \citet{beccari17} SFH would look if sampled with a constant log age step of 0.2 dex.
    \label{fig:hist_becc}
    }
\end{figure*}

To test the statistical model described in Sec.~\ref{sec:inference} we apply it to the simulated Orion A populations described above. We consider four versions of our statistical model to show the effects of making no corrections, correcting for UBS, correcting for extinction, and finally correcting for both binaries and extinction, the latter producing our best estimates of the SFH, since it exploits most aspects of the data. These four versions are listed in incremental order of complexity in Table~\ref{tab:tmodel}.

In Fig.~\ref{fig:hist_all} and Fig.\,\ref{fig:hist_becc} we compare the SFH inferred from our Orion A and ONC simulations in Sec.~\ref{app:oriona} using the S, S+E and S+UB+E models (Table~\ref{tab:tmodel}). We show the results when we include
{\it (a)} all the stars in the simulation ($G_{lim}$\,=\,$\infty$), and
{\it (b)} only the stars brighter than $G_{lim}$\,=\,8.5 mag.
The {\it blue dots} and {\it blue error bars} indicate, respectively, the median $\Psi$ and the credibility interval (Sec-\,\ref{sec:nummass}) resulting from the Bayesian inference solution for the set of 10 simulations for $G_{lim}$\,=\,$\infty$. The {\it red dots} and {\it red error bars} correspond to the case $G_{lim}$\,=\,8.5. The results for the S+E and S+UB+E models have been corrected for incompleteness as described in Sec.\,\ref{app:incompcorr}.

The black dots in Fig.\,\ref{fig:hist_all} are drawn at the value $\Psi$\,=\,200$\,{\rm M}_\odot\,{\rm Myr}^{-1}$ used in the simulation. We analyzed 10 individual realizations of 4000 stars for each set of isochrones. For the case shown in the {\it top panels} of Fig.~\ref{fig:hist_all} we repeated the simulations of Sec.~\ref{app:oriona} but this time including only single stars with no extinction by dust, i.e., the same assumptions of the S model. In this case the simulated stars track exactly the isochrones in the CMD, which explains the excellent match between the inferred and the {\it true} ({\it black dots}) SFHs.

For the {\it middle} and {\it bottom panels} of Fig.~\ref{fig:hist_all} we use the simulations shown in Fig.~\ref{fig:CMDtt}. We see from the figure that the agreement between the inferred and {\it true} SFHs deteriorates as we include more parameters in the statistical model. This is natural because both extinction by dust and UBS introduce a {\it blurring effect} on the sample, resulting on a more uncertain age assigned statistically to each star by our model. Notwithstanding these facts, the agreement between the inferred and {\it true} SFHs is satisfactory.

The {\it black} lines in Fig.\,\ref{fig:hist_becc} show 
the bursty SFH described in section \ref{app:oriona}. We analyzed 10 different realizations of this catalog. The {\it left} panel shows the SFH derived from a simulation which includes only single stars and no extinction. Our solution fits the assumed SFH because in the absence of extinction and UBS, our procedure matches  the age of the simulated stars with low uncertainty. The {\it middle} and {\it right} panels show the mean SFH inferred using the S+E and the S+UB+E models, respectively. In the {\it middle} panel, the contributions at $\log({\rm Age}/{\rm yr})=5.8$ and $6.4$ dex are biased, shifting the mean towards younger ages. We expect this shift because the S+E model ignores the UBS present in the simulations. The S+UB+E model, on the other hand, corrects statistically for the presence of  UBS, and the inferred SFH ({\it right} panel) does not show this bias, the SFRs are slightly overestimated, but fit the
\citet{beccari17} values within the credibility intervals.

We conclude that whereas we cannot recover exactly the three bursts proposed by 
\citet{beccari17}, our solution reproduces the overall distribution of age suggested in their study. It is challenging to discern individual peaks of SF from the summed up distribution shown in \citet{beccari17} figure 6 and table 1. Moreover, disentangling these peaks becomes complex when the effects of dust are significant and the fraction of UBS is high. Our results from Sec.\,\ref{sec:results} imply that the width and strong asymmetry of the 
\citet{beccari17} distribution is inconsistent with what we obtain analyzing Gaia data.

With these caveats in mind, these tests show that we recover the {\it true} SFH used in the simulations with enough accuracy and precision as to consider our statistical model a reliable tool to infer the SFH of Orion A.

\subsection{Tests on real data}

\begin{figure*}
\begin{center}
    \includegraphics[width=\textwidth]{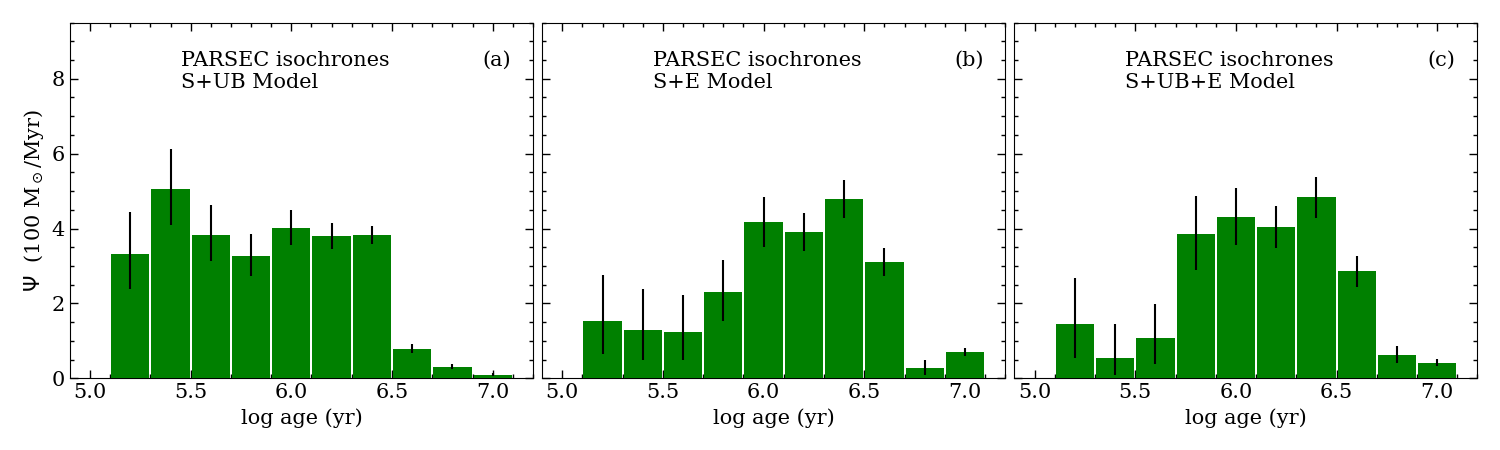}
    \caption{SFH inferred from the \citetalias{mck19} sample of 1275 stars using the PARSEC isochrones. In all cases $G_{\rm lim}$\,=\,16.5 mag.
    The characteristics of the different models are described in Table~\ref{tab:tmodel}.
    The results in panels {\it (b,c)} have been corrected for sample incompleteness (Sec.~\ref{app:incompcorr}).
    Model S+UB+E is to be preferred since it exploits most aspects of the data.
    \label{fig:sfh_results3x1parsec}}
\end{center}
\end{figure*}

Fig.~\ref{fig:sfh_results3x1parsec} shows the inferred Orion A SFH using the PARSEC isochrones with the S+UB, S+E and S+UB+E statistical models (Table~\ref{tab:tmodel}). For this derivation of the SFH we limit the \citetalias{mck19} sample to stars with apparent $G$ magnitude brighter than $G_{\rm lim}$\,=\,16.5 mag. The values of $\epsilon_{8.5}$ listed in columns 5 to 7 of Table~\ref{tab:imass} are used to correct the SFH inferred from the S+E and S+UB+E models. Direct comparison of panels {\it (b)} and {\it (a)} of Fig.\,\ref{fig:sfh_results3x1parsec}, shows that ignoring extinction yields a large number of stars with $t$(yr)\,<\,6 dex: extinction moves stars towards the region occupied by young objects (protostars) in the CMD, and heavily reddened stars are then assigned young ages by the S and S+UB Models. The solutions for the S+E and S+UB+E models are qualitatively very similar, indicating that modelling extinction is far more important than accounting for the presence of UBS. This is because the magnitude of the reddening vector exceeds by a large amount the brightening effect of UBS. The direction of these vectors in the CMD are also quite different, introducing different biases in the uncorrected samples.
The correction for UBS is thus small in comparison with the correction for extinction. The same behaviour is observed in Fig.~\ref{fig:sfh_results_b}. 

For clarity and for the benefit of the reader, in the main body of this paper we show only results for the S+UB+E model and omit referring to this model by this name (cf. Fig.~\ref{fig:sfh_results}).

\label{lastpage}

\end{document}